%% file: Draft_YG_22Oct_Final.tex
\documentclass[draftcls, 12pt,onecolumn]{IEEEtran}
\input{preamble3.tex}
\allowdisplaybreaks[1]
\usepackage{float}
\usepackage{cite}
\title{Wireless Compressive Sensing for Energy Harvesting Sensor Nodes}

\author{Gang~Yang, Vincent Y.~F.~Tan, Chin~Keong~Ho, See~Ho~Ting, and Yong~Liang~Guan  
\thanks{G. Yang, S. H. Ting and Y. L. Guan are with the School of Electrical and Electronic Engineering, Nanyang Technological University, Singapore (e-mail:yang0305@e.ntu.edu.sg; \{shting, eylguan\}@ntu.edu.sg). G.~Yang is supported in part by the Advanced Communications Research Program DSOCL06271, a research grant from the Directorate of Research and Technology (DRTech), Ministry of Defence, Singapore. } 
\thanks{V. Y. F. Tan and C. K. Ho are with the Institute for Infocomm Research, A$^\star$STAR, Singapore (e-mail: \{tanyfv, hock\}@i2r.a-star.edu.sg). V. Y. F. Tan is also with the Department of Electrical and Computer Engineering, National University of Singapore.}
}

\begin{document}
\maketitle 

\vspace{-1in}
\begin{abstract}
We consider the scenario in which multiple sensors send spatially correlated data to a fusion center (FC) via independent Rayleigh-fading channels with additive noise. Assuming that the sensor data is sparse in some basis, we show that the recovery of this sparse signal can be formulated as a compressive sensing (CS) problem. To model the scenario in which the  sensors operate with intermittently available energy that is harvested from the environment, we propose that each sensor transmits independently with some probability, and adapts the transmit power to its harvested energy. Due to  the  probabilistic transmissions, the elements of the equivalent sensing matrix are not Gaussian. Besides, since the sensors have different energy harvesting rates and different sensor-to-FC distances, the FC has different receive signal-to-noise ratios (SNRs) for each sensor. This is referred to as the {\em inhomogeneity} of SNRs. Thus, the elements of the sensing matrix are also not identically distributed. For this unconventional setting, we provide theoretical guarantees on the number of measurements for reliable and computationally efficient  recovery, by showing that the sensing matrix satisfies the restricted isometry property (RIP), under reasonable conditions. We then compute an achievable  system delay under an allowable mean-squared-error (MSE). Furthermore, using techniques from large deviations theory, we analyze the impact of inhomogeneity of SNRs on the so-called $k$-restricted eigenvalues, which governs the number of measurements required for the RIP to hold. We conclude that   the number of measurements required for the RIP is not sensitive  to the inhomogeneity of SNRs, when  the number of  sensors $n$ is large and the sparsity of the sensor data (signal) $k$ grows slower than the square root of $n$.  Our analysis is corroborated by extensive numerical results.
\end{abstract}


\section{Introduction}
The lifetimes of conventional wireless sensor networks (WSNs) are limited by the total energy available in the batteries. It is inconvenient to replace batteries periodically, or even impossible when sensors are deployed in harsh conditions, e.g., in toxic environments or inside human bodies. Energy harvesting of ambient energy such as solar, wind, thermal and piezoelectric energy, appears as a promising alternative to a fixed-energy battery, to prolong the lifetime and offer potentially maintenance-free operation for WSNs\cite{Kansal,CKTSP2012}. Compared to limited but reliable power supply from conventional batteries, energy harvesters provide a virtually perpetual but unreliable energy source. Moreover, the sensors typically have different {\emph{energy harvesting rates}}, due to varying harvesting conditions such as the spread of sunlight and difference in wind speeds.

This paper addresses the problem of data transmission in energy harvesting WSNs (EH-WSNs). We assume that energy harvesting sensors are deployed to monitor some physical phenomenon in space, e.g., temperature, toxicity of gas. Data collected from sensors are sent to the fusion center (FC). The data are typically correlated, and well approximated by a sparse vector in an appropriate transform (e.g., the Fourier transform). Recent developments in compressive sensing (CS) theory provide efficient methods to recover sparse signals from limited measurements\cite{IntroductionCS06}. CS theory states that if the sensing matrix satisfies the  restricted isometry property (RIP), a small number  of measurements (relative to the length of the data vector) is sufficient to accurately recover the sparse data. This advantage of CS potentially allows us to reduce the total number of transmissions, and this is particularly important for data transmission in bandwidth-limited wireless channels.

The accurate estimation of the sensor data by the FC has recently been addressed  by using CS techniques  in the literature. In \cite{ReconRandProjHaupt2006}, Haupt {\emph{et. al}} presented a sensing scheme based on phase-coherent transmissions for all sensors. However, \cite{ReconRandProjHaupt2006} made two practically limiting assumptions. First, it  assumed that there was no channel fading, and path losses for all sensors were identical. Second, the transmissions from all sensors were synchronized such that signals arrived in phase at the FC. In \cite{InfBoundAeron2007}, Aeron {\emph{et. al}} derived information theoretic bounds on sensing capacity of sensor networks under a fixed signal-to-noise ratio (SNR) for all sensors. In contrast, \cite{EASTRana2010} proposed a sparse approximation method in non-fading channels, which adapted a sensor's sensing activity according to its energy availability. In \cite{XueICC2012}, Xue {\emph{et. al}} successively applied CS in the spatial domain and the time domain, under a fixed SNR for all sensors. In \cite{Fazel2011}, Fazel {\emph{et. al}} proposed a random access scheme in underwater sensor networks. Each activated sensor picked a uniformly-distributed delay to transmit. By simply discarding the colliding data packets from concurrent medium access, the FC used a CS decoder to recover the sensor data based only on the successfully received packets. Thus, the scheme did not exploit packet collisions for data recovery.

Since sensors are placed at different locations, it is commonly assumed that the sensors transmit data over independent but nonidentical channels with different fading conditions. Different energy harvesting rates also lead to different transmit powers and hence different (receive) SNRs. We refer to this generally as the {\em inhomogeneity} of SNRs. The application of {\em wireless compressive sensing} to the scenario of inhomogeneous SNRs has, to the best of our knowledge, not been studied in the literature. We define the {\emph{system delay}} as the number of concurrent sensor-to-FC transmissions (or channel uses) for estimating one data vector (among sensors). We aim to reduce the system delay, while ensuring a target estimation accuracy. Surprisingly, we observe that the required number of measurements for accurate recovery $m$ is not overly  sensitive to the inhomogeneity of SNRs provided that the number of sensors  $n$ is large and the sparsity of the data vector $k$ grows slower than $\sqrt{n}$. This motivates us to further investigate the impact of inhomogeneity of SNRs, based on the recovery performance in terms of RIP.

The three main contributions are summarized as follows.

\begin{enumerate}
  \item We first present an efficient transmission scheme, which features probabilistic transmission by sensor nodes. In each time slot, every sensor locally decides to transmit with some probability, and adjusts the transmit power according to its energy availability. The FC thus receives a linear combination of signals that are transmitted from a random subset of sensors. The transmissions over successive time slots result in a sensing matrix which is effectively achieved through the mixing of signals in wireless channels.
  \item Second, we prove that the FC can recover the data accurately, if the total number of transmissions (or measurements) \(m\) exceeds 
  \begin{align}
    O \left( k \frac{\rho_{\max} (k)}{\rho_{\min} (k)} \log \frac{n}{k} \right),
  \end{align}
  where $n$ is the number of sensors, \(k\) is the sparsity of the sensor data, and \(\rho_{\max} (k) \) and \( \rho_{\min} (k)\) are respectively the maximum and minimum \(k\)-restricted eigenvalues (see definition in \eqref{eq:eq410}) of a Gram matrix which depend on the inhomogeneity of SNRs. Different from previous works on CS, our bound depends explicitly on the ratio  \(\rho_{\max} (k) / \rho_{\min} (k)\), which is the \(k\)-restricted condition number of the Gram matrix.   Based on this result, we also compute the achievable  system delay subject to a desired recovery accuracy.
  \item Third, we analyze the impact of inhomogeneity of SNRs on the required number of measurements, in terms of $\rho_{\max}(k)$ and $\rho_{\min}(k)$. We model the signal powers of the sensors as independent truncated Gaussians. By using the theory of large deviations, we show that both $\rho_{\max}(k)$ and $\rho_{\min}(k)$ concentrate around one (for all constant $k$) in large $n$ regime, and the rate of convergence to one depends on the inhomogeneity of SNRs.
      This allows us to explain the   observation that the inhomogeneity of SNRs does not significantly affect the number of measurements required for the RIP to hold.
\end{enumerate}

This remainder of this paper is organized as follows: Section \ref{sec2:system-model} provides a description of the system model.  Section \ref{sec3:energy-aware-WCS} presents a new wireless compressive sensing scheme. Section \ref{SecIV} details the main results on the RIP, the achievable system delay and investigates the impact of inhomogeneity of SNRs. Section \ref{SecV} provides the simulation results. Section \ref{SecVI} concludes this paper. The proofs for the RIP result and the result on the impact of inhomogeneity of SNRs  are given in Section \ref{SecVII}. 

We adopt the following set of notation in this paper: lower case letters denotes deterministic scalars, and lower case Greek letters for constants or angles. Boldface upper case and boldface lower case refer to matrices and (column) vectors, respectively. We use upper case letters to denote random variables. Sets are denoted with calligraphic font (e.g., \(\calV\)). The cardinality of a finite set \(\calV\) is denoted as \(|\calV|\). The $n$-order identity matrix is denoted by $\bI_n$. We also use \(\bbR^n\) and \(\bbC^n\) to denote the \(n\)-dimensional real and complex Euclidean spaces respectively. 

\section{System Model} \label{sec2:system-model}
Consider a wireless sensor network that consists of \(n\) energy harvesting sensor nodes and a FC. Sensors transmit their  data to the FC via a shared multiple-access channel (MAC). We consider slotted transmissions by first considering a single snapshot of the spatial-temporal field. Assuming the sensor data \(\bs\) is compressible, we can model it as being sparse with respect to (w.r.t.) to a fixed orthonormal basis $\{ \bpsi_j  \in \bbC^n : j=1,\ldots, n\}$, i.e.,
\begin{equation}\label{eq:basis decomposition}
\bs = \bPsi \bx = \sum_{j=1}^n \bpsi_j x_j,
\end{equation}
where \(\bx\in\bbC^n\) has at most $k  < \lfloor n/2 \rfloor$ non-zero components and $\lfloor \cdot \rfloor$ is the floor operation.

We assume a flat-fading channel with complex-valued channel coefficients $h_{ij}$, where $1\leq i \leq m$ denotes the slot index and $ 1\leq j\leq n$ denotes the sensor index. The channel remains constant in each slot. We further assume a Rayleigh-fading channel, hence the channel coefficients for different slots are independent and identically distributed (\iid) according to the complex Gaussian distribution.

We propose that sensors concurrently transmit to the FC in a probabilistic manner, such that the signals from sensors are linearly combined over the air. Sensor \(j\) multiplies its datum \(s_j\) by some random amplitude \(\phi_{ij}\) (to be defined in \eqref{eq:eq31}), then transmits in the \(i\)-th time slot. The FC thus  receives
\[y_i  = \sum_{j {=} 1}^n {h_{ij} {\phi_{ij}} s_j } + e_i,
\]
where \(e_i\) is a noise term (not necessarily Gaussian). After \(m\) time slots, the FC receives the measurement vector
\begin{equation} \label{eq:model0}
\begin{array}{l}
{\by} = ({\bH} \odot {\bPhi} ){\bs} + {\be} = \bZ \bs + {\be} =\bZ \bPsi \bx  + {\be}  ,
\end{array}
\end{equation}
where the matrix \(\bZ = {\bH} \odot {\bPhi}\), and the operation \( \odot \) is  the element-wise product of two matrices. We assume all noise components are independent, zero mean and have variance \(\sigma^2\). The signal model over one slot is illustrated in Fig. \ref{fig:Fig1}.

\input{mac2}

From the perspective of signal recovery, we want  to estimate \(\bx\) or equivalently \(\bs\), from \({\by}\), such that the {\em mean-squared-error} (MSE)   $\bbE \|{ \hatbx} - \bx\|_2^2$   does not exceed some threshold $\epsilon$. Also, we would like to estimate the sparse vector using  minimum network resources (i.e., channel uses), due to limited channel resources. Thus, given a fixed number of sensors \(n\) and an $\epsilon$, our objective is to design a transmission scheme that minimizes the number of sensor-to-FC transmissions \(m\).

Different from \cite{EASTRana2010,PhDThesisBajwa2009}, we consider Rayleigh-fading channels, and adopt concurrent transmissions in a probabilistic manner. Moreover, the SNRs of different sensors are considered to be different, compared to the fixed SNR case in the literature \cite{InfBoundAeron2007,XueICC2012,PhDThesisBajwa2009}.

\section{Energy-Aware Wireless Compressive Sensing} \label{sec3:energy-aware-WCS}
In Section~\ref{sec3a:CS-prospective}, we first provide a CS perspective for the signal model in \eqref{eq:model0}. Then in Section ~\ref{sec3b:energy-aware-trans-strategy}, we present an energy-aware wireless compressive sensing scheme. By taking into account the inhomogeneity of SNRs, we also derive the probability distribution function (pdf) of elements in the random matrix \( \bZ\) in Section~\ref{sec3c:pdf-analysis}, which will be used to show the RIP in Section~\ref{Section4A}.

\subsection{A Compressive Sensing Perspective}\label{sec3a:CS-prospective}
Since we assume the data vector $\bx$ is sparse in some basis, it seems natural to adopt a CS method to recover \(\bx\). The over-the-air combination via the channel matrix $\bH$ contributes to the effective equivalent sensing matrix \(\bZ\) in \eqref{eq:model0}. However, there are two differences from the conventional CS setup that make the analysis more challenging.
\begin{itemize}
\item Due to probabilistic transmissions, the elements of the sensing matrix $\bZ$ are not Gaussian.
\item Since sensors have different energy harvesting rates and different sensor-to-FC distances, the FC has different receive SNRs for all sensors. Thus, the elements of the sensing matrix $\bZ$ are also not identically distributed.
\end{itemize}
The proposed transmission scheme calls for the analysis of non-Gaussian non-\iid \ sensing matrices. Hence, we need to analyze the system performance in a more intricate way that differs from conventional CS problems. The key technique we employ is to show that the elements of the sensing matrix $\bZ$ are sub-Gaussian, and make use of new results on sub-Gaussian random matrices.

\subsection{Energy-Aware Wireless Compressive Sensing}\label{sec3b:energy-aware-trans-strategy}
We consider only the energy consumption for wireless transmissions, by assuming the energy consumption on sensing is negligible. The energy harvesting rate  varies over sensors. For simplicity, we assume that each sensor allocates the same power for all slots. Let $E_j$ be the accumulated harvested energy that is available for sensor $j$ to transmit in each slot. We perform energy-aware wireless transmissions taking into account the different available energy. It is noted that a {\em causal energy constraint} that comes from energy harvesting should be satisfied, i.e., energy that is consumed for transmissions can not exceed the energy available in each slot.

Set a probability \(p \in (0,1]\) and a squared-amplitude $b_j > 0$ . Let \({\bPhi}\) in \eqref{eq:model0}  be a {\em selection-and-weight} (SW) matrix, whose elements are independently generated according to the   random variable
\begin{equation} \label{eq:eq31}
\phi _{ij}  =  \left\{ \begin{array}{cl}
  + \sqrt{b_j} & \mbox{w.p. } p/2 \\
  0 & \mbox{w.p. }1 - p,  \quad \;\; \forall \; i =1,2,\dots,m.\\
  - \sqrt{b_j} & \mbox{w.p. } p/2 \\
 \end{array} \right.
\end{equation}
That is, the sensor \(j\) transmits with probability \(p\) with an amplitude of \(\sqrt{b_j}\), and the actual value is positive or negative with equal probability. Given available energy $E_j$, we choose $b_j$ such that\footnote{The quantity \(b_{j}\) can be written more generally as \(b_{i,j}\), which means the transmit powers for different slots are different. To reduce the complexity of processing, we allocate the same power to all the slots.}
\begin{align}
  p b_j \leq E_j,  \quad \forall j=1,2,\ldots,n. \label{CausalEnergyConstraint}
\end{align}
Clearly, each entry \(\phi _{ij}\) is   zero mean and has variance \(p b_j\). The causal energy constraint is satisfied in expectation, i.e.,  \( \bbE (\phi_{ij}^2 ) =p b_j \leq E_j\). This allows us to save energy to be used for future transmissions. The energy-saving feature can be crucial in the scenario where the energy harvesting rates are fluctuating over several snapshots of the spatial-temporal field. It is, however, beyond the scope of this paper to optimize for the $b_j$'s.

In \cite{EASTRana2010}, all sensors consume  the same amount of energy for transmissions. In contrast, each sensor here adapts the transmit power to its available energy via the above-designed SW matrix. Furthermore, the SW matrix randomly selects the sensors to transmit, and weighs the data according to the sensors' harvested energy. In each time slot, a subset of sensors are selected at random to perform transmissions and over-the-air combination. The selections are performed in a distributed manner at each sensor node, since each node separately decides the slots that it transmits in. We couple random sensor selection and energy-aware transmission by the choice of the  SW matrix.

Recall the signal model in \eqref{eq:model0}, i.e., ${\by} = \bZ \bPsi \bx  + {\be}$. With the knowledge{\footnote{The assumption that the FC knows \(\bZ\) and \(\bPsi\) is reasonable, because the FC can perform channel estimation from preambles, and obtain the information on the amount of harvested energy via feedback. The channel and energy information is used for generating SW matrix from a predefined set of SW matrices. The global parameters like \(m\) and \( p\) can be broadcasted to all sensors.}} of the matrix \(\bZ\) and the sparsity-inducing basis \(\bPsi\), the FC can implement   CS decoding to recover sparse coefficients \(\hatbx\) and obtain the estimated data vector \(\hatbs = \bPsi \hatbx\).

\subsection{Probability Distribution Analysis and Equivalent Normalized Signal Model} \label{sec3c:pdf-analysis}
Consider the signal model in (\ref{eq:model0}). Denote each element in \(\bZ\) as \(Z_{ij} = h_{ij} \phi_{ij} = Z_{ij}^{\mathrm{R}}+j Z_{ij}^{\mathrm{I}} \), where \(Z_{ij}^{\mathrm{R}} \triangleq h_{ij}^{\mathrm{R}} \phi_{ij},\) and \( Z_{ij}^{\mathrm{I}} \triangleq h_{ij}^{\mathrm{I}} \phi_{ij}\).  Note that elements of the matrix \({\bH}\) are assumed to be independent, and each element $h_{ij}$ has independent real and imaginary components. Also the matrix \(\bPhi\) consists of independent elements. All elements of matrix \(\bZ\) are thus independent, and have independent real and imaginary components. As such, it suffices to analyze the probability distribution of the real component, since the analysis is similar for the imaginary component. The marginal pdf of \( Z_{ij}^{\mathrm{R}} \) can be shown to be
\begin{align} \label{eq:eq33}
 f_{Z_{ij}^{\mathrm{R}}} (z) &= \frac{1}{\sqrt{b_j}} f_{H_j^{\mathrm{R}}} \left(\frac{z}{\sqrt{b_j}} \right) \cdot \frac{{p}}{2} + \frac{1}{\sqrt{b_j}} f_{H_j^{\mathrm{R}}} \left(- \frac{z}{\sqrt{b_j}} \right) \cdot \frac{{p}}{2}\; + (1-p) \cdot {{\delta}(z)}, 
\end{align}
where \(f_{H_j^{\mathrm{R}}}(\cdot)\) is the pdf of channel coefficient of sensor \(j\), and \(\delta(\cdot)\) is the Dirac delta function. For the sake of brevity, we define a new pdf as follows.

\begin{mydef} \label{mydef:def1a}
A random variable \(X\) follows a {\emph{mixed Gaussian}} distribution, denoted as \(X \sim {{\widetilde{{\calN}}}} (\mu, \nu^2, p)\), if its pdf has the following form
\begin{equation} \label{eq:eq34}
 f_X(x) =p \frac{1}{\sqrt{2\pi \nu^2}} \exp{\left(- \frac{(x- \mu)^2} {2 \nu^2}\right)}  + (1-p) {{\delta}(x)},
\end{equation}
where \(p \in (0,1]\) is the mixing parameter. The corresponding complex mixed Gaussian distribution, assuming the real and imaginary components are independent, is denoted as \({{\widetilde{{\calN}}}}_c (\mu, \nu^2, p)\).
\end{mydef}

Assuming Rayleigh-fading channels, all elements in the channel matrix \( \bH \) are independent, zero mean and follow Gaussian distributions. Note that due to different fading channels for the sensors, the matrix \( \bH \) has column-dependent variances, where the $j$-th column follows a Gaussian distribution  with variances \(\nu^2_j\). From \eqref{eq:eq33} and \eqref{eq:eq34}, the marginal pdf of \( Z_{ij}^{\mathrm{R}}\) can be expressed as
\begin{align}\label{eq:eq35}
 f_{Z^{\mathrm{R}}} (z)
  &=p \frac{1}{\sqrt{ \pi   \nu_j^2 b_j }} \exp{\left(- \frac{z^2}{\nu_j^2 b_j}\right)}
  + (1-p) {{\delta}(z)}.
\end{align}
Thus, we have \(Z^{\mathrm{R}} \sim {\widetilde{{\calN}}} \big(0,\nu_j^2 b_j /2, p \big) \).

Recall that $\bZ ={\bH} \odot \bPhi $. Let $\bH = {\tilbH} {\bGamma}_{\rmH}$ and $\bPhi={\tbPhi} {\bGamma}_{\Phi}$, where \({\bGamma_{\rmH}} = {\diag} \lbrace \nu_1, \nu_2, \ldots,\) \(\nu_n \rbrace \) and \({\bGamma_{\Phi}} = {\diag} \lbrace \sqrt{p b_1}, \sqrt{p b_2},\) \(\ldots, \sqrt{p b_n} \rbrace\). Then we can decompose the matrix \(\bZ\) as follows
\begin{equation} \label{eq:eq36}
{\bZ} = \sqrt{m} {\tilbZ} {\bGamma},
\end{equation}
where we denote $\tilbZ ={\tilbH} \odot \tbPhi $ and   \({\bGamma}={\bGamma}_{\rmH} {\bGamma}_{\Phi}\). Let \({\bGamma} = {\diag} \lbrace \sqrt{\gamma_1}, \sqrt{\gamma_2}, \ldots, \sqrt{\gamma_n} \rbrace \), where the receive signal power of sensor \(j\) is{\footnote{The receive signal power depends on both the channel condition (i.e., the variance of fading coefficients $\nu_j^2$  and the average transmit power $p b_j$) that is governed by the accumulated harvested energy.}} \(\gamma_j = p b_j \nu_j^2\). We term the diagonal elements of $\bGamma$ a signal power pattern. The $\gamma_j$'s are generally  all different (i.e., inhomogeneous signal powers), and this directly leads to the inhomogeneous (receive) SNRs. We note that all elements of the matrix \(\tilbZ\) are \iid \ mixed Gaussian random variables, i.e., \({\tilZ} \sim {\widetilde{{\calN}}}_c \left(0,1/(pm), p\right)\) and \({\tilZ}^{\mathrm{R}} \sim {\widetilde{{\calN}}} \left(0,1/{(2pm)}, p\right)\).

Using the equivalent expression in (\ref{eq:eq36}), we rewrite the signal model in \eqref{eq:model0} as
\begin{equation} \label{eq:eq46b}
\by = \sqrt{m} {\tilbZ} \bGamma \bPsi \bx + \be,
\end{equation}
where the matrix \(\bPsi\) is a unitary matrix. The distinct signal powers in $\bGamma$ are spread along sparsity-inducing basis vectors (i.e., columns of \(\bPsi\)).

A matrix (or more correctly, a {\em sequence} of matrices) is said to be {\em standard column regular} if all elements are uniformly bounded by some constant\cite{RMTinWC2004}. For analytical convenience, we normalize the matrix \(\bGamma \bPsi\) to be standard column regular. The normalization constant is  $ \| \bGamma \bPsi \|_F / \sqrt{n}   = \sqrt{{P}_{\rm{ave}}}$, where \({ P}_{\rm{ave}} = \sum \nolimits_{j=1}^n p b_j \nu_j^2 / n\) denotes the average (receive) signal power in one time slot.   Then the normalized matrix 
\begin{equation} \label{eq:eq38}
 \bSigma = \bGamma \bPsi / \sqrt{{ P}_{\rm{ave}}}
\end{equation}
has bounded spectral norm. By dividing both sides of (\ref{eq:eq46b}) by \(\sqrt{m { P}_{\rm ave}}\), we    obtain the normalized signal model
\begin{equation} \label{eq:eq39}
\tilby = \tilbZ \bSigma \bx + \tilbe = \bA \bx + \tilbe,
\end{equation}
where all noise components are independent,   zero mean and have normalized variance \(\tsigma^2 \triangleq  \sigma^2 / ( m {P}_{\rm{ave}}) \). The average SNR is defined as
\begin{align}
  {\rm{SNR}}_{\rm{ave}} \triangleq  \frac{{{P}}_{\rm{ave}}} {\sigma^2} = \frac{p}{n \sigma^2} \sum_{j=1}^n b_j \nu_j^2. \label{AverageSNR}
\end{align}

\section{Main Results} \label{SecIV}
Having derived the probability distribution of elements of the matrix \(\bZ\) in Section~\ref{sec3c:pdf-analysis}, we recall  the definition of RIP~\cite{{DLPCandes05}}  and state our main result, that is Theorem \ref{mythe:the1}, in Section \ref{Section4A}. The engineering implication of Theorem \ref{mythe:the1}, and in particular the tradeoff between the achievable system delay and the allowable MSE, will be discussed in~\ref{Section4B}. Finally we analyze the effect of inhomogeneity of SNRs on RIP and the required number of measurements in Section~\ref{Section4C}.

\subsection{Restricted Isometry Property}\label{Section4A}
It is well established in CS theory that  a sufficient condition for accurate and efficient recovery (via convex optimization) is that the sensing matrix satisfies the RIP. A   matrix \(\bA\) is said to satisfy RIP of order \(k\), if there exists a \(\delta_k \in (0,1)\) such that
\begin{equation} \label{eqn:RIP}
(1 - \delta_k) {\norm22{\bx}} \le \left\| {\bA \bx} \right\|_2^2 \le (1 + \delta_k) \norm22{\bx}
\end{equation}
holds for all \(k\)-sparse vectors $\bx$. The smallest constant $\delta_k$ satisfying \eqref{eqn:RIP} is known as the {\em restricted isometry constant} (RIC)~\cite{{DLPCandes05}}. When the sensing matrix \(\bA\) is random, the inequality should hold with overwhelming probability that approaches one as \(n\) grows. Many families of random matrices, e.g., \iid \ Gaussian random matrices and Bernoulli random matrices are known to satisfy the RIP\cite{DLPCandes05,SimpleRIP2008}. As a result, to evaluate the recovery performance, all we have to show is that the sensing matrix \(\bA\) in our scheme also obeys RIP with overwhelming probability.

The RIP requires that the sensing matrix \(\bA\) preserves the Euclidean norm of sparse vectors well. For the signal model in \eqref{eq:eq39}, the entries in \(\tilbZ\) are \iid \ {\em sub-Gaussian} random variables (See Definition \ref{def5a} in Section \ref{SecVIIA}). It is known that random matrices (with sufficiently many rows and) with \iid \ sub-Gaussian entries approximately preserve the Euclidean norm of sparse vectors with high probability~\cite{CSEldar2011}. Since \(\bA = \tilbZ \bSigma\), we need to analyze the norm-preserving property of \(\bSigma\). To do so,
we define the {\emph{k-restricted extreme eigenvalues}} of the Gram matrix \({\bf {\Sigma^{\ast} \Sigma}}\) as
\begin{align}\label{eq:eq410}
\begin{split}
  \rho_{\max}(k) &=\max_{ \bv: \|\bv\|_0\le k , \|\bv\|_2= 1 }  { \| \bSigma \bv \|_2^2}, \\
\rho_{\min}(k) &=\min_{ \bv: \|\bv\|_0\le k , \|\bv\|_2= 1 }  { \| \bSigma \bv \|_2^2},
\end{split}
\end{align}
where $\bv \in \bbC^n$, and the ``\(l_0\)-norm'' \(\| \bv \|_0\) refers to the number of non-zero elements of $\bv$. The extreme eigenvalues will be used to understand how the inhomogeneous SNRs affects the RIP.

\begin{mylem}\label{lem:lem1}
The following bounds on $\rho_{\max}(k)$ and  $\rho_{\min}(k)$  hold:
\begin{equation} \label{eqn:bounds_evalues}
1\le\rho_{\max}(k)\le k,\qquad 0\le\rho_{\min}(k)\le 1.
\end{equation}
\end{mylem}

\begin{IEEEproof}
Fix a vector $\bv \in \bbC^n$ such that $\|\bv\|_2=1$ and $\|\bv\|_0=k$. Let $\calT \subset\{1,\ldots, n\}$ with $|\calT|\le k$ be the support of $\bv$. Let  $\bSigma_{\calT} \in\bbC^{n\times |\calT|}$ be the submatrix of $\bSigma$ with column indices $\calT$. Denote the eigenvalues of the Gram matrix $\bSigma_{\calT}^* \bSigma_{\calT}$ by $\lambda_1 \ge\ldots\ge\lambda_k\ge 0$. Due to the normalization in \eqref{eq:eq38}, the trace of $\bSigma_{\calT}^* \bSigma_{\calT}$ is $\sum_{j=1}^k\lambda_j=k$. This implies that the largest eigenvalue is at least one and at most $k$. Similarly, the smallest eigenvalue is no larger than one.
\end{IEEEproof}

We note that the sparsity level \(k\) is usually much smaller than the number of sensors \(n\) in  large-scale WSNs. We further assume \(\rho_{\max}(k) \in [1,2]\) in the following. This simplifies some of the mathematical arguments. We   analytically and numerically verify this claim in Section \ref{Section4C}.  To state our main theoretical result cleanly, we define two quantities that depend on \(\bSigma\) and \(k\) as follows
\begin{equation} \label{eq:eq410b}
\begin{split}
\xi_k (\bSigma) &\triangleq \max \left\{ 1-\rho_{\min}(k), \rho_{\max}(k) -1\right\}, \\
\zeta_k (\bSigma) &\triangleq \max \left\{ 0, \frac{2-\rho_{\max}(k)-\rho_{\min}(k)} {\rho_{\max}(k)-\rho_{\min}(k)} \right\}.
\end{split}
\end{equation}
Since $\rho_{\max}(k)\in [1,2]$, we have\footnote{The arguments of some quantities are sometimes omitted for notational convenience.} \(\xi_k, \zeta_k \in [0,1]\). Let $\vartheta_k=(1+\zeta_k) \rho_{\max}(k) -1$. Given \(\delta_k \in (\xi_k, 1)\), for convenience, we map \(\delta_k\) to a ``modified RIC'' via a piecewise linear mapping as follows
\begin{equation} \label{eq:eq411}
\beta_k (\delta_k, \bSigma) \triangleq \left\{ \begin{array}{l}
 1 - (1 - \delta_k) / {\rho_{\min}(k)},  \quad \delta_k \in ( \xi_k,  \vartheta_k ) \\
(1 + \delta_k)/{\rho_{\max}(k)} -1, \quad  \delta_k \in ( \vartheta_k, 1 ) .
\end{array} \right.
\end{equation}
Let $\varsigma_k=2 / \rho_{\max}(k) -1$. The inverse of $\beta_k(\delta_k,\bSigma)$ is denoted as
\begin{equation} \label{eq:eq412}
\begin{array}{l}
\delta_k (\beta_k, \bSigma) \triangleq  \left\{ \begin{array}{l}
1 - (1- \beta_k) \rho_{\min}(k),  \quad\, \beta_k \in (0, \zeta_k )  \\
(1+ \beta_k) \rho_{\max}(k) -1,  \quad\beta_k \in ( \zeta_k, \varsigma_k ).
 \end{array} \right.
\end{array}
\end{equation}
In the sequel, we assume that the quantity \(\xi_k (\bSigma)\) is a small positive number and it measures the inhomogeneity of the eigenvalues of $\bSigma_{\calT}^*\bSigma_{\calT}$ for $|\calT|\le k$. This implies \(\zeta_k  \) is small, and the deviation between \(\beta_k\) and \(\delta_k\) is also small. The validity of this assumption will be shown both analytically and numerically in Section \ref{Section4C}.

Recall that the sensing matrix \(\bA = \tilbZ \bSigma\) in \eqref{eq:eq39}, where all elements of the \(m \times n\) matrix \(\tilbZ\) are \iid \ mixed Gaussian random variables,  $\bSigma$ is defined in \eqref{eq:eq38},  and $n$ is the number of sensors.  We  now state our main theoretical result.

\newpage
\begin{mythe} \label{mythe:the1}
Let $c_1,c_2>0$ be some universal constants. Given a sparsity level \( k < \lfloor n/2 \rfloor\), a transmit probability \(p \in (0,1]\) and a number \( \delta_k \in (\xi_k, 1)\), if the number of measurements satisfies
\begin{equation} \label{eq:eq413}
m > \frac{c_1 k \rho_{\max}(k)} {p^2 \beta_k^2 {\rho_{\min}(k)}} \log{\frac{5en}{k}},
\end{equation}
where $\beta_k= \beta_k (\delta_k, \bSigma)$ is defined in \eqref{eq:eq411}, then for any vector \(\bx\) with support of cardinality of at most \(k\), we have  that the RIP in~\eqref{eqn:RIP} holds with probability at least
\begin{equation} \label{eq:eq415}
1- \exp{\left(-  {{c_2} m p^2 \beta_k^2} / {4} \right)} .
\end{equation}
\end{mythe}

\begin{IEEEproof}
  See Section \ref{SecVIIA}.
\end{IEEEproof}


\begin{myrem} [Specialization to the homogeneous case] \label{myrem:rem1}
Clearly, the lower bound on the required number of measurements is \(O (\frac{k \rho_{\max}(k)} {\beta_k^2 {\rho_{\min}(k)}} \log{\frac{n}{k}})\). For the homogeneous signal power pattern (i.e., the matrix \(\bGamma\) is a multiple of the identity matrix $\bI_{n}$), we have \(\rho_{\max}(k) = \rho_{\min}(k) = 1 \) and \(\beta_k = \delta_k\). Thus the lower bound reduces to \(O (\frac{k} {\delta_k^2} \log{\frac{n}{k}})\), which coincides with the known results for \iid \ random sensing matrices. See Theorem 5.2 in \cite{SimpleRIP2008} and Section 1.4.4 in \cite{CSEldar2011}.
\end{myrem}

\begin{myrem} [Contribution to the RIP analysis] \label{myrem:rem2}
Due to the inhomogeneous signal power pattern, the rows \(\ba_i\) of the  sub-Gaussian sensing matrix \(\bA\) are {\em non-isotropic}. To the best of our knowledge, little is known about the RIP of non-isotropic sub-Gaussian random matrices. The only relevant result is in  Remark \(5.40\) in \cite{CSEldar2011} which gives a concentration inequality of non-isotropic random sensing matrices in terms of the upper bound on the spectral norm. However, the authors did not demonstrate how the inhomogeneity affects the RIP, nor did they investigate the number of measurements required to satisfy the RIP. Theorem \ref{mythe:the1} fills  this gap. 
\end{myrem}

\begin{myrem}\label{myrem:rem3}
Theorem \ref{mythe:the1} is proved in Section \ref{SecVIIA} by leveraging Theorem 2.1 of \cite{UUPSubgaussian2008}, which states that a sufficient condition for the approximate preservation of the  Euclidean norm upon random linear mapping is that the number of measurements is proportional to the fourth power of the sub-Gaussian norm. In our scenario, as shown in Lemma \ref{lem5},  the sub-Gaussian norm bounded above by $1/\sqrt{p}$. In addition, Lemma \ref{lem5} shows  (using the Chernoff-bound)  that the sub-Gaussian tail probability is  bounded above by $pe^{-p t^2/2}$. Note that the sub-Gaussian norm is the smallest constant $\varrho > 0$ for which the sub-Gaussian tail  probability is $2e^{- t^2/(2 \varrho^2)}$ (Definition~\ref{def5a}). In view of the fact that the pre-factor in our bound is $p$ (and not $2$), there is some degradation with respect to $p$ in Theorem 1. For larger $p$, the degradation is reduced. 
\end{myrem}

\subsection{Achievable System Delay}\label{Section4B}
The performance of wireless compressive sensing scheme is characterized by two quantities, i.e., the MSE and the system delay. The MSE performance under bounded noise is studied in the CS literature\cite{IntroductionCS06,NewRICCai,DavenportNoisefold11}. Note that there is often a trade-off between the two quantities. Under an allowable MSE  \(\epsilon > 0\), we thus analyze the {\em achievable  system delay} $D(\epsilon)$, which is defined as
\begin{equation} \label{eq:eq419e}
  D(\epsilon) \triangleq \min_{m} \,\,\, m \;\;\; \st \;\;\; \bbE \|\hatbx -\bx\|_2^2 \leq \epsilon.
\end{equation}

\begin{mycor} \label{mycor:cor1}
Let \(p, m, n, k, \bSigma, \xi_k, \vartheta_k\) be as in Theorem \ref{mythe:the1}. Let $\epsilon_{\mathrm{th}}\triangleq 1/(0.0942 \times {\rm SNR}_{\rm{ave}})$. Given an allowable MSE  \(\epsilon >\epsilon_{\mathrm{th}}\), with overwhelming probability (exceeding~\eqref{eq:eq415}), the achievable system delay is
\begin{equation} \label{eq:eq419f}
  D(\epsilon) = \frac{c_1 k \rho_{\max}(k) } {p^2 (\tilde{\beta_k})^2 {\rho_{\min}(k)}} \log{\frac{5en}{k}},
\end{equation}
where
\begin{equation} \label{eq:eq419f2}
\tilde{\beta_k} (\bSigma, \epsilon) \triangleq \left\{ \begin{split}
& 1 - \frac{0.693 + 1 / \sqrt{ \epsilon {\rm SNR}_{\rm{ave}}} } {\rho_{\min}(k)},  \quad \delta_k \in ( \xi_k, \vartheta_k ), \\
& \frac{1.307 - 1 / \sqrt{ \epsilon {\rm SNR}_{\rm{ave}} }} {\rho_{\max}(k)} -1, \quad   \delta_k \in  ( \vartheta_k, 1 ).
\end{split} \right.
\end{equation}
\end{mycor}

\begin{IEEEproof}
We start the proof by leveraging on the following lemma.
\begin{mylem}[Theorem 3.2 of \cite{NewRICCai}] \label{mylem:pro}
Let \(\tilby = {\bA \bx} + \tilbe\), where \(\bx\) is a $k$-sparse vector in \(\bbC^n\), \(\tile \in \bbC^m\) is a zero mean, white random vector whose entries have variance \(\sigma^2 \). If the \(\bA\) satisfies the RIP with RIC \(\delta_k < 0.307\), then the solution \(\hatbx\) to the $\ell_1$-minimization problem in CS decoder \cite{IntroductionCS06,CSEldar2011} satisfies
\begin{equation} \label{MSEBound}
\bbE \|{ \hatbx} - \bx\|_2^2  \leq  \frac{\sigma^2} { { P}_{\rm{ave}} (0.307- \delta_k)^2 }.
\end{equation}
\end{mylem}

Recall the definition of \({\mathrm{SNR}_{\rm{ave}}}\) in \eqref{AverageSNR}. From Lemma \ref{mylem:pro}, to achieve a MSE  \(\epsilon\), it suffices to ensure the RIC satisfies $\delta_k^{\ast} = 0.307 - 1 / \sqrt{\epsilon {\rm SNR}_{\rm{ave}}}$. From Theorem \ref{mythe:the1}, the required minimum number of measurements  such that the RIP holds with overwhelming probability is
\begin{align}
  m_{\min} = \frac{c_1 k \rho_{\max}(k) \log{\frac{5en}{k}}}{p^2 (\tilde{\beta_k^{\ast}})^2 {\rho_{\min}(k)}}
\end{align}
where $\tilde{\beta_k}$ is given in \eqref{eq:eq419f2}. The definition of the achievable system delay   establishes Corollary~\ref{mycor:cor1}.
\end{IEEEproof}

\begin{myrem} \label{myrem:rem10}
Note that Corollary \ref{mycor:cor1} applies only to the case where the MSE $\epsilon$ is greater than the threshold $\epsilon_{\mathrm{th}}$. If $\epsilon \leq \epsilon_{\mathrm{th}}$, then from \eqref{MSEBound}, simple algebra reveals that $\delta_k=0$, which implies that the sensing matrix $\bA$ is a perfect isometry. Since $\bA$ is random, and the entries are governed by a density that is absolutely continuous w.r.t.\ the Lebesgue measure, this occurs with probability zero, implying that the constraint in \eqref{eq:eq419e} is almost surely not satisfied. Thus,  in this case, we define the system delay to be $\infty$.
\end{myrem}

\begin{myrem} \label{myrem:rem4}
As either $\epsilon$  or ${\mathrm{SNR}}_{\rm{ave}}$ increases, $\tilde{\beta_k}$ increases, and thus the system delay $D(\epsilon)$ decreases. More importantly, we note from Corollary \ref{mycor:cor1} that the key measure for the inhomogeneity of SNRs is the ratio \(r(k) \triangleq \rho_{\max} (k) / \rho_{\min} (k) \in [1,\infty)\). The system delay increases as $r(k)$ increases from one. We hence analyze the impact of inhomogeneity of SNRs on the deviation of $\rho_{\max} (k)$ and $ \rho_{\min} (k)$ from unity in Section \ref{Section4C}. In addition, the system delay decreases as $p$ increases, since ${\mathrm{SNR}}_{\rm{ave}}$ defined in \eqref{AverageSNR} increases as $p$ increases. Thus, there is an inherent tradeoff between system delay and energy consumption  because large $p$ implies high transmit energy. Thus, it is always advantageous to transmit with as  high a probability as possible subject to the causal energy constraint.
\end{myrem}

\begin{figure}[t]
\centering
\includegraphics[width=.65\columnwidth] {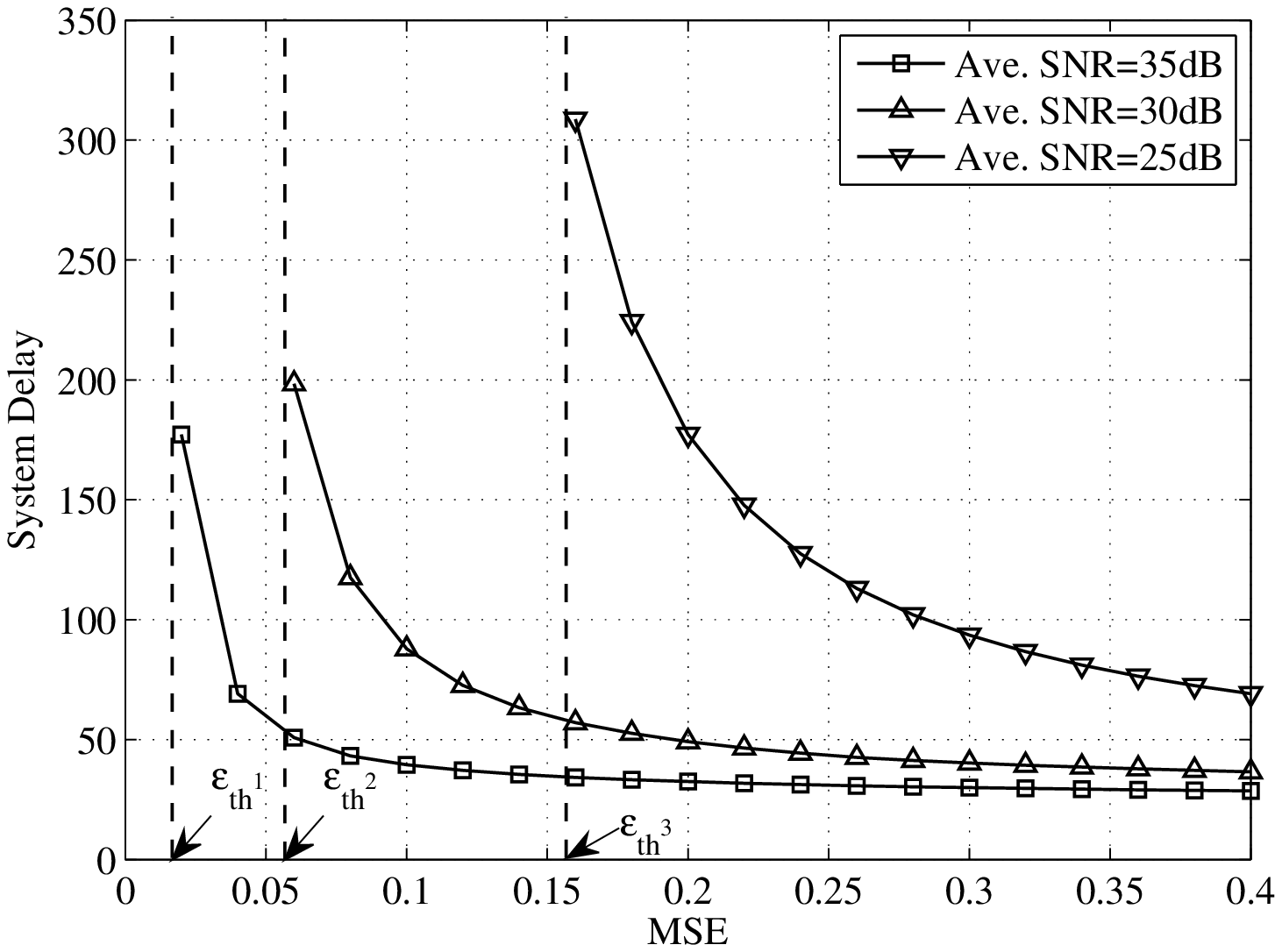}
\caption{Plot of achievable system delay against allowable MSE. Beyond the MSE threshold, the achievable system delay increases as either the allowable MSE or the average SNR increases.}
\label{fig:Fig2}
\end{figure}

\begin{myexa}
Let the number of sensors \(n=500\), the sparsity level \(k=5\) and the  transmit probability $p=0.8$. These parameters imply $\rho_{\max}(k)=1.09, \rho_{\min}(k)=0.88$ (See Section~\ref{SecIV}).  We plot the achievable system delay $D(\epsilon)$ against the allowable MSE $\epsilon$, for different average SNRs in Fig. \ref{fig:Fig2}. We observe that beyond the MSE threshold (that depends on the average SNR), the system delay $D(\epsilon)$ decreases as either $\epsilon$ or ${\mathrm{SNR}}_{\rm{ave}}$ increases, which is is expected.
\end{myexa}

\begin{myrem} \label{myrem:rem4b}
We considered the scenario in which the FC collects one data vector from all sensors in one frame. As a generalization of our setup, one can seek to minimize the total number of slots for collecting multiple data vectors. By adjusting the transmit probability in each frame, one can allocate different powers for different frames, such that both the recovery accuracy and the causal energy constraint is guaranteed. Details of this possible extension are beyond the scope of this paper.
\end{myrem}

\subsection{Effect of Inhomogeneity}\label{Section4C}
This section investigates the impact of inhomogeneity of (receive) SNRs on the number of measurements needed to satisfy the RIP. Without loss of generality, we assume all sensors have the same noise power, hence, it suffices to analyze the impact of inhomogeneity of receive signal powers. We focus on the asymptotic scenario  where the number of sensors $n$ tends to infinity and, for the ease of analysis, $k$ is kept constant. To make the dependence on $n$ clear, we denote $\rho_{\max}(k)$ (resp.\ $\rho_{\min}(k)$) as $\rho_{\max}(k,n)$  (resp.\ $\rho_{\min}(k,n)$). It will be shown that both $\rho_{\max}(k,n)$ and $\rho_{\min}(k,n)$ concentrate around one when $n$ is large, and the rate of convergence to one depends on the inhomogeneity of SNRs. This implies that the recovery performance (the required number of measurements and the probability that the RIP holds in Theorem \ref{mythe:the1}) is not sensitive to the inhomogeneity of SNRs when $n$ is large.

Let \({\bw} = \bSigma \bv \), where the unit-norm, \(k\)-sparse vector \(\bv\) is supported on the set \(\calT \triangleq \{s_1,   \dots, s_k\}\subset\{1,\ldots, n\}\) and let \(s_1 < \ldots < s_k\). To obtain further insights, we let   \(\bPsi\) be the \(n\)-point discrete Fourier transform (DFT) matrix. Then the  squared $\ell_2$-norm of $\bw$ can be expressed as follows
\begin{equation} \label{eq:eq420}
\| {\bw} \|_2^2 = \frac{1}{n { P}_{\rm{ave}}} \sum_{i=1}^n \gamma_i  \left(  1 +  \sum_{q=1}^k  \sum_{l=1, l < q}^k  2 \rmR \rme \left\{ { v}_{s_q} { v}_{s_l}^{\ast} \exp \left({\frac{- j 2 \pi (i-1) (s_q - s_l) } {n} } \right) \right\}  \right).
\end{equation}
Since $\|\bw\|_2^2$ is strongly influenced by the inner summation terms, we analyze the  behavior of these terms   more carefully in the sequel. When the signal power pattern is homogeneous, i.e., \({\bGamma} = { \diag (\sqrt{\gamma},   \ldots, \sqrt{\gamma})}\), we have \( \| {\bw} \|_2^2 = \| \bSigma \bv \|_2^2 = 1 \), hence  \(\rho_{\max}(k,n) = \rho_{\min}(k,n) = 1 \) for all $k,n$.

We are interested to know how $\rho_{\max}(k,n)$ and $\rho_{\min}(k,n)$ vary with different signal powers $\gamma_i$'s. Thus, we consider a model in which the $\gamma_i$'s are i.i.d.\ random variables following an approximate Gaussian distribution. By varying the variance of this distribution, we are in fact varying the inhomogeneity of the signal powers. Specifically, to deal with the fact that the signal powers cannot be negative, we use the following truncated Gaussian distribution to model the signal powers. 

\begin{mydef}\label{mydef:def2}
A random variable $X$ is {\em truncated Gaussian}, denoted as $\calN_{\mathrm{tr}}(\mu, \omega^2)$, if its pdf is
\begin{equation} \label{eq:eq421b}
g_X(x ; \mu, \omega)= \frac{1} {\sqrt{2 \pi} \omega (1-Q(\mu / \omega))} \exp \left(- \frac{(x-\mu)^2}{2 \omega^2}\right),
\end{equation}
for $x \geq 0$ and $0$ else, where $Q(x)\triangleq \frac{1}{\sqrt{2 \pi}} \int_{x}^{\infty} e^{- {t^2}/{2} }  \, dt$ is the $Q$-function of a standard Gaussian pdf.
\end{mydef}

We assume that $\gamma_i \sim \calN_{\mathrm{tr}}(\mu, \omega^2)$ for all $i=1 ,\dots,n$ and they are mutually independent. Given $\mu$, the ``variance'' $\omega^2$ is   a measure of the degree of inhomogeneity of the signal powers $\gamma_i$'s. Also, the parameter $d\triangleq\mu/\omega$ is a measure of the homogeneity of the SNRs. If $d$ is small (resp.\ large), the SNRs are less (resp.\ more) homogeneous. We use the exponential asymptotic notation $a_n\dotleq \exp(-nE)$ to mean that $\limsup_{n\to\infty}\frac{1}{n}\log a_n\le -E$.  Under the above assumptions on the statistics of the signal powers, we have the following large deviations upper bound on $\rho_{\max}(k,n)$ and $\rho_{\min}(k,n)$:
\begin{mythe} \label{mythe:the2}
Let  $d\triangleq \mu / \omega$. For any $t >0$,  and any constant $1 \leq k < \lfloor n/2 \rfloor$, 
\begin{equation} \label{eq:eq423}
\begin{split}
\bbP \left( \rho_{\max} (k,n) > 1+ t \right) \dotleq \exp \left[- n  d^2 E (k, t)^2   \right], \\
\bbP \left( \rho_{\min} (k,n) < 1- t \right) \dotleq  \exp \left[- n  d^2 E (k, t)^2  \right],
\end{split}
\end{equation}
where  the exponent $E (k, t)$  is defined as $E (k, t) \triangleq t /  (k-1+\sqrt{2} t)$.
\end{mythe}

\begin{IEEEproof}
  See Section \ref{SecVIIB}.
\end{IEEEproof}

Recall  that Theorem \ref{mythe:the1} says that both the required number of measurements and the probability that the RIP holds depends on the ratio $r(k,n)=\rho_{\max}(k,n)/\rho_{\min}(k,n)$. From Theorem \ref{mythe:the2}, we note that both $\rho_{\max}(k,n)$ and $\rho_{\min}(k,n)$ concentrate around one in the large $n$ regime (for  bounded $k$), and the rate of convergence to one depends on the inhomogeneity of SNRs. This allows us to conclude that that for large-scale EHWSNs (relative to the signal sparsity), the inhomogeneity of SNRs does not significantly affect the RIP and the system delay, which is a surprisingly positive observation.

\begin{myrem}\label{myrem:rem5}
We note that $E (k, t)$ is an increasing function of $t$ and a decreasing function of the sparsity $k$ which is   expected.  Also, the exponent $d^2E (k, t)^2$ increases with $d$, which means that the convergence of  $\rho_{\max} (k,n)$ and $ \rho_{\min} (k,n)$ to unity is faster when $d$ is large, or equivalently, when the signal powers are more homogeneous. It is observed that $\rho_{\max}(k,n)$ is close to one in the large $n$ regime. This validates the assumption that $\rho_{\max}(k,n) \in [1,2]$ in Section \ref{Section4A}.
\end{myrem}

\begin{myrem}\label{myrem:rem7}
In the preceding analysis, and particularly in Theorem~\ref{mythe:the2}, we assumed that $k$ does not grow with $n$. Close examination of the proof shows that if $k= \lfloor n^{1/2-\lambda} \rfloor $ for any $\lambda\in (0,1/2]$, then the probability that   $\{\rho_{\max}(k,n)>1+t\}$   still  goes  to zero albeit at a slower rate  of $\approx \exp(-   n^{2\lambda} d^2 t^2 )$ (not exponential in $n$). More precisely, we can verify that
\begin{equation}
\limsup_{n\to\infty}\frac{1}{n^{2\lambda}}\log \bbP( \rho_{\max}(k,n)>1+t )\le  -d^2 t^2, \label{eqn:md}
\end{equation}
and analogously for   $\{\rho_{\min}(k,n)<1-t\}$. Inequality~\eqref{eqn:md} is a so-called {\em moderate-deviations}  result \cite[Sec.\ 3.7]{LargeDeviation1998}. Notice that the dependencies on the homogeneity $d=\mu/\omega$ and $t$ are similar to \eqref{eq:eq423}.
\end{myrem}

\begin{myrem}\label{myrem:rem6}
One may wonder whether Theorem~\ref{mythe:the2} depends strongly on $\bPsi$ being the DFT matrix. In fact, the only property of the DFT that we exploit in the proof of Theorem \ref{mythe:the2} is its circular symmetry, i.e., each basis vector of the DFT (containing elements that are powers of the $n$-th root of unity) is uniformly distributed over the circle in the complex plane. Hence, certain Ces\`aro-sums converge to zero and the proof goes through. See~\eqref{eqn:cesaro} in Section \ref{SecVIIB}. Thus, Theorem \ref{mythe:the2} also applies for other sparsity-inducing bases whose basis vectors have the  circular  symmetric property, e.g., the discrete cosine transform (DCT) or the Hadamard transform. 
\end{myrem}

\section{Simulation Results}\label{SecV}
We now numerically validate our results.  We set the number of sensors \(n=500\)  and transmit probability \(p=0.8\). We use the truncated Gaussian distribution with  $\mu=0.2$ to model the receive signal powers, and use the basis pursuit de-noising (BPDN) algorithm \cite{BPDNBerg} as the CS decoder.

\begin{figure}[t] 
\centering
\includegraphics[width=.65\columnwidth] {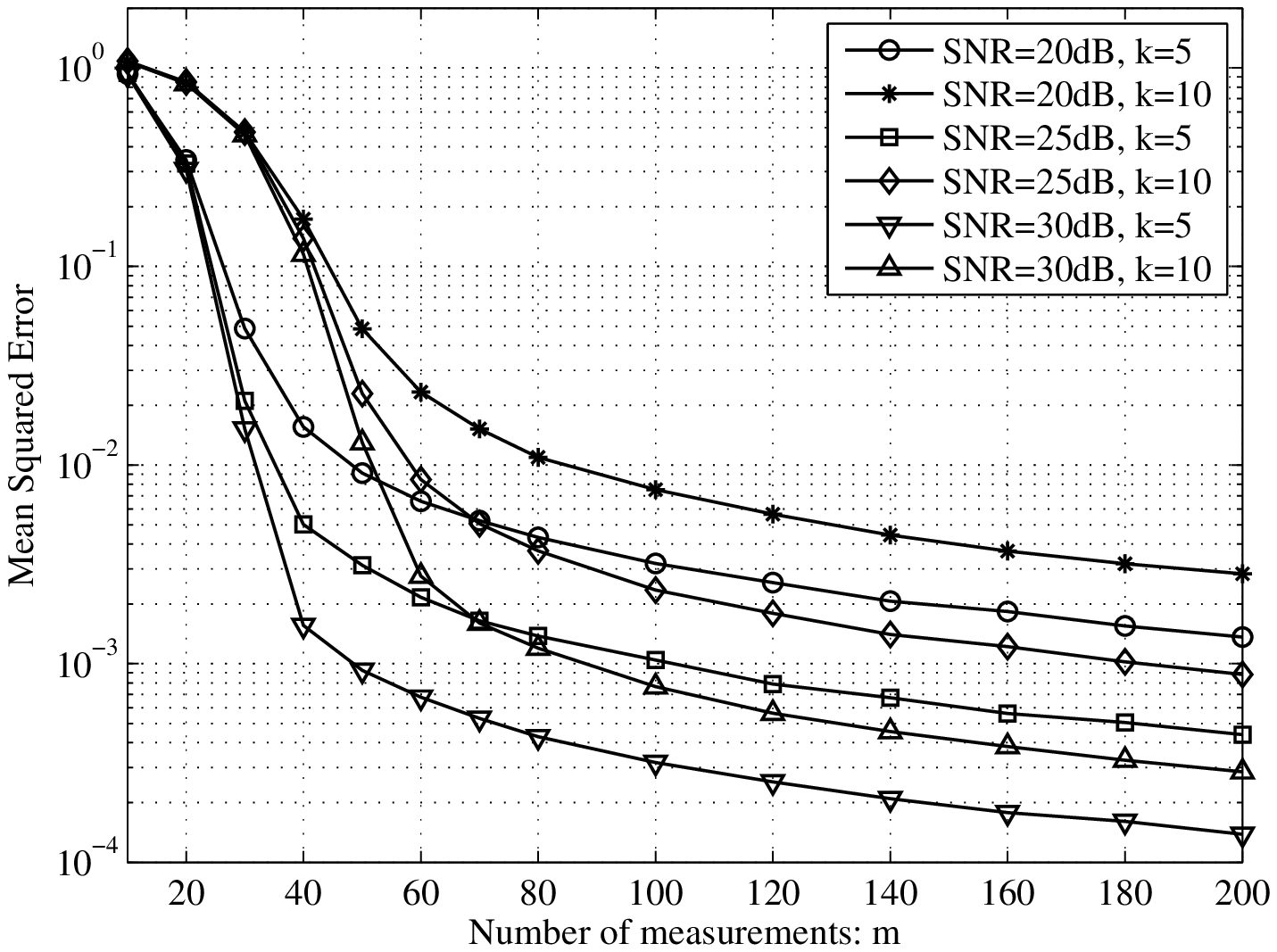}
\caption{Plot of MSE against the number of measurements. The MSE decreases as \(k\) decreases, or the average SNR increases.}
\label{fig:Fig3}
\end{figure}

First, we fix $d=2$, which implies $\omega=\mu/d=0.1$. Fig. \ref{fig:Fig3} plots the MSE against the number of measurements (or transmissions) \(m\) for different sparsities \(k\) and different average SNRs. As expected, the MSE decreases as either \(k\) decreases  or the average SNR increases. Consider the MSE level \(2 \times 10^{-3}\). When the average SNR is $25$ dB, the wireless compressive sensing scheme achieves a smaller system delay of \(D = 68\) for \(k=5\) compared to  \(D = 115\) for \(k=10\). When the sparsity $k=5$, the scheme achieves a smaller system delay of \(D = 39\) for ${\rm{SNR}_{\rm{ave}}}=30{\rm dB}$ compared to  \(D = 68\) for ${\rm{SNR}_{\rm{ave}}}=25 {\rm dB}$.

\begin{figure}[t] 
\centering
\includegraphics[width=.65\columnwidth] {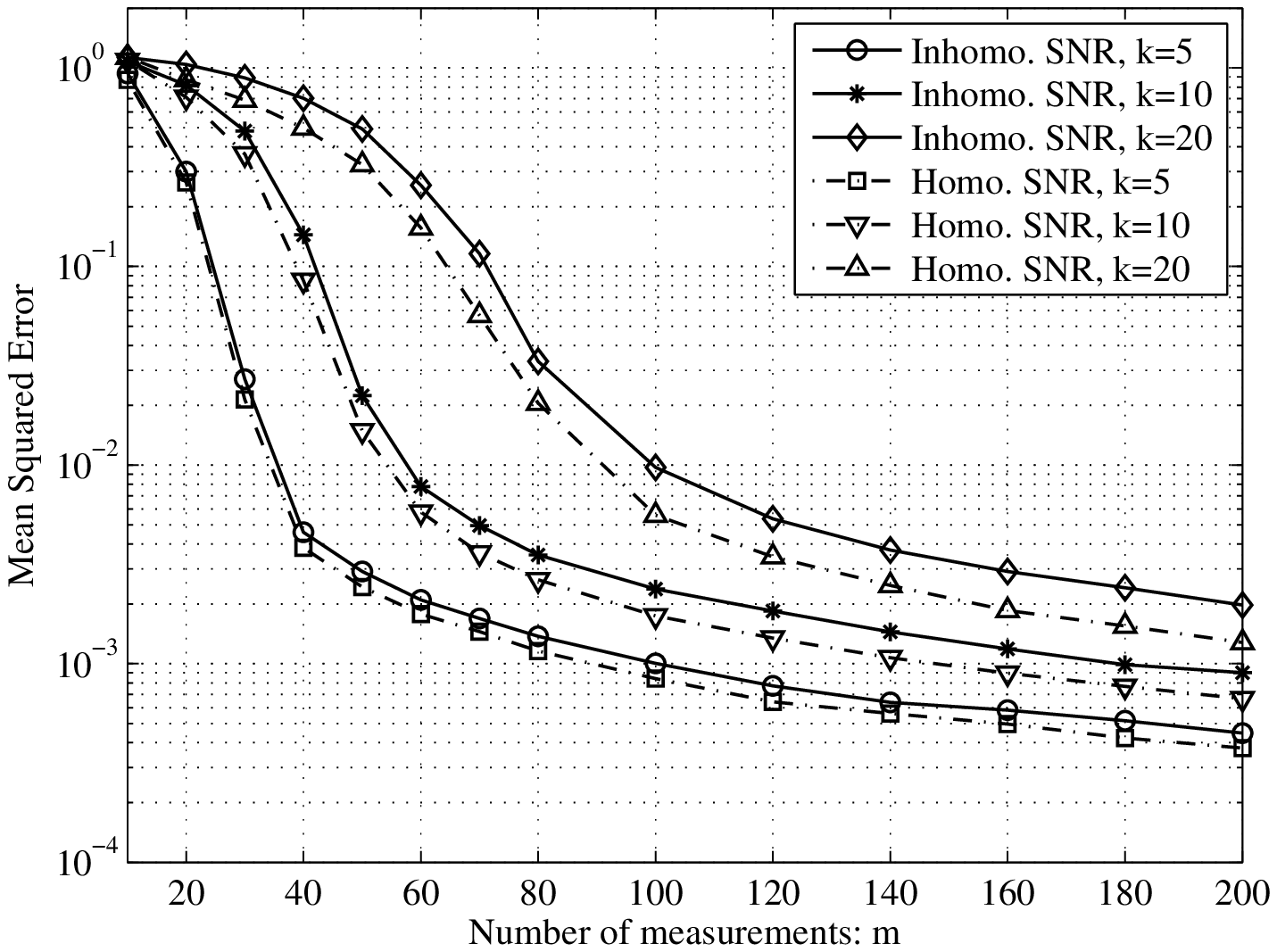}
\caption{Plot of MSE against the number of measurements. The MSE performance for the inhomogeneous scenario is slightly worse than that of the homogeneous-SNR scenario. }
\label{fig:Fig4}
\end{figure}

Second, we fix $d=2$ and the average SNR to be $25 {\rm dB}$. Fig. \ref{fig:Fig4} compares the MSEs of the inhomogeneous SNR  and the homogeneous SNR scenarios, for the sparsity levels $k=5, 10, 20$. It is observed that in the inhomogeneous scenario, the MSE performance is slightly worse than that of the homogeneous-SNR scenario. Note that the degradation becomes larger  as the sparsity $k$ increases. This is because the convergence rate for   $\rho_{\max}(k)$ and $\rho_{\min}(k)$ to one is faster  if $k$ is small relative to $n$. This corroborates the observation in Section \ref{Section4C}.


\begin{figure}[t] 
\centering
\includegraphics[width=.65\columnwidth] {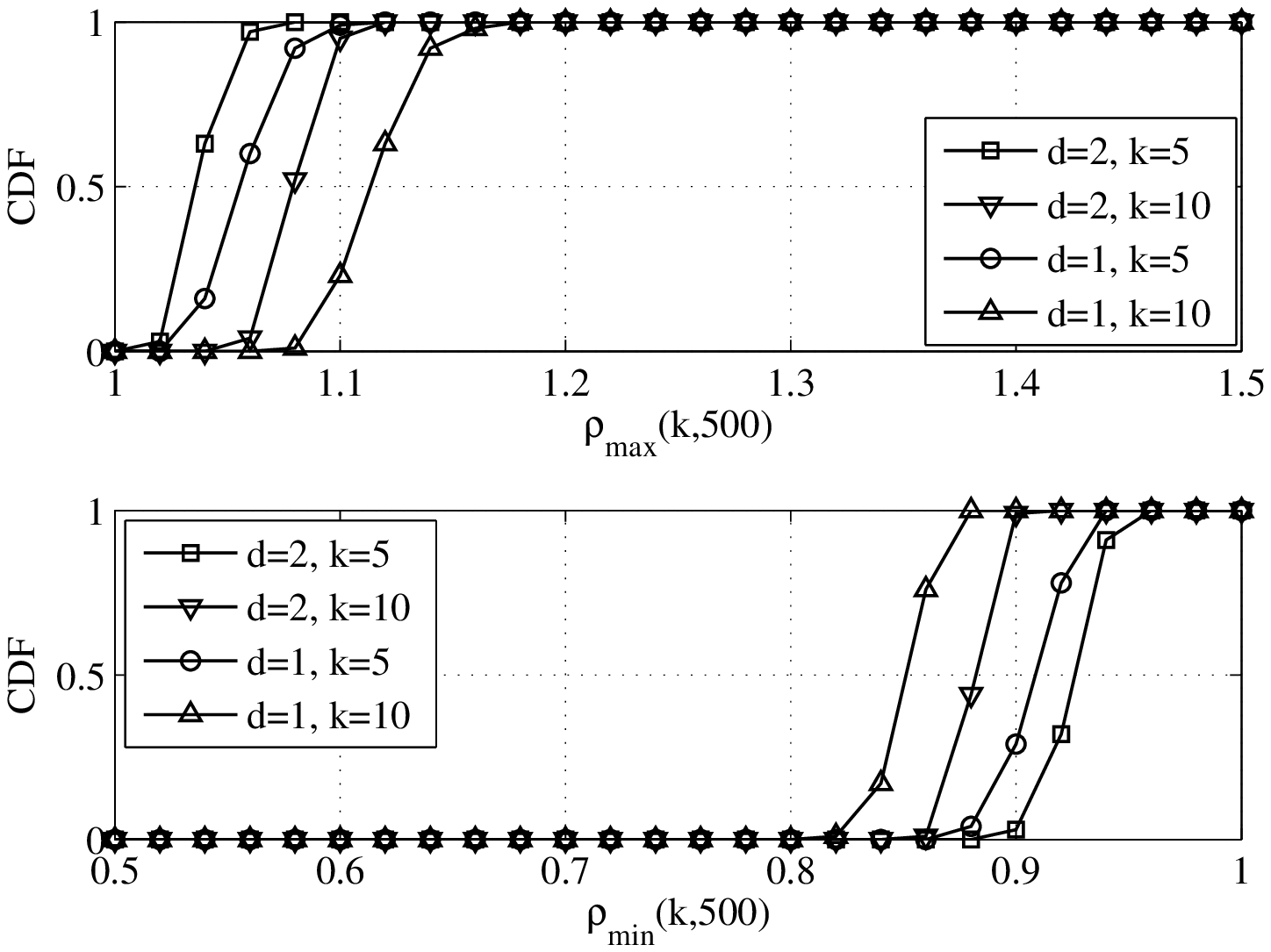}
\caption{CDF of \(\rho_{\max}(k,500)\) and \( \rho_{\min} (k,500) \). Both converge to one faster for more homogeneous SNRs (i.e., large $d$).}
\label{fig:Fig5}
\end{figure}

Third, we set $d=1,2$ and $k=5, 10$. Fig. \ref{fig:Fig5} shows the cumulative distribution function (CDF) of $\rho_{\max}(k,500)$ and $\rho_{\min}(k,500)$. We note that both $\rho_{\max} (k,500)$ and $ \rho_{\min} (k,500)$ converge to one faster for larger $d$, or equivalently, for more homogeneous SNRs. Also, under the same inhomogeneous SNRs, both $\rho_{\max} (k,500)$ and $ \rho_{\min} (k,500)$ converge to one faster for smaller $k$.

\begin{figure}[t] 
\centering
\includegraphics[width=.65\columnwidth] {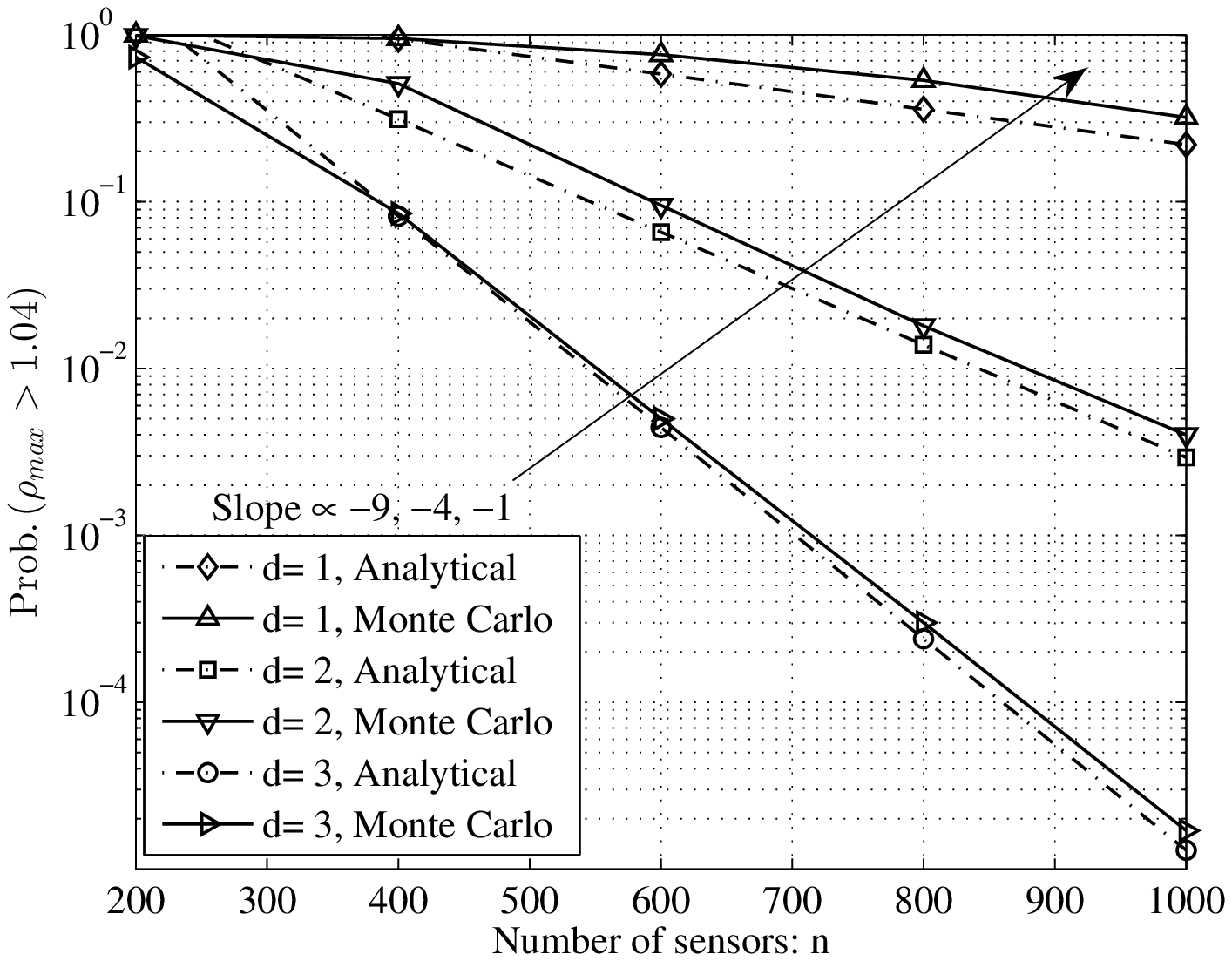}
\caption{Plot of the probability of $\rho_{\max}(k,n)>1.04$ against the number of sensors. The logarithm of the probability decreases linearly as $n$ grows, and the slope varies quadratically w.r.t. $d$.}
\label{fig:Fig6}
\end{figure}

Finally, we numerically validate the asymptotic behavior of $\rho_{\max}(k,n)$ as $n$ grows. Set $k=5$, $d=1, 2, 3$, respectively. Fig.\ \ref{fig:Fig6} shows the probability that $\rho_{\max}(k,n)>1.04$ for different $n$. It is observed that the logarithm of the probability decreases linearly as $n$ grows (when $n/k$ is large) and furthermore, the slope varies quadratically w.r.t. $d$, i.e., the slope is proportional to $-1, -4,-9$ for $d=1, 2, 3$, respectively. This observation corroborates Theorem \ref{mythe:the2}. 

\section{Conclusion}\label{SecVI}
In this paper, we considered the scenario in which each sensor independently decides whether or not to transmit with some probability $p$, and  the overall transmission power (and thus $p$) depends on its available energy. Hence, only a subset of sensors transmits concurrently to the FC, and this exploits the spatial combination inherent in  wireless channels. We use techniques from CS theory to prove a lower bound on the required number of measurements to satisfy the RIP and hence to ensure that the data recovery is both computationally efficient (and amenable to convex optimization) and accurate. We also compute an achievable system delay given an allowable MSE. Finally, we analyze the impact of inhomogeneity on the $k$-restricted extreme eigenvalues. These eigenvalues govern the number of measurements required for the RIP to hold. In large-scale EH-WSNs,  we showed using large deviation techniques that the recovery accuracy and the system delay are not sensitive to the inhomogeneity of SNRs. 

\section{Proofs of Main Results} \label{SecVII}
\subsection{Proof of Theorem \ref{mythe:the1}} \label{SecVIIA}
\begin{IEEEproof}
Recall the signal model in \eqref{eq:eq39}, i.e., \(\tilby = \tilbZ  \bSigma \bx + \tilbe = {\bA \bx} + \tilbe \). The proof involves three steps. In step 1 and step 2, we prove the desired result when all quantities are real; and in step 3, we extend the result to the complex case. For the real case, we show that the matrix \(\tilbZ\) acts as isometry on the images of the sparse vector under matrix \(\bSigma\), i.e., on the set \(\{\bSigma \bv: \|\bv\|_0 \leq k , \bv \in \bbR^n\}\). By showing the rows of \(\tilbZ\) are isotropic sub-Gaussian and by exploiting the so-called ``restricted eigenvalue property'' of \(\bSigma\), we derive an RIP for the matrix \(\bA\) in step 2. Before step 1, we start with the following preliminaries. Let \( d(\bu,\bv)\) be the Euclidean distance in \(\bbR^n\). 


\begin{mydef}[Nets, covering numbers \cite{CSEldar2011}] \label{def1}
Consider a metric space \((\calU, d)\) with \(\calU \subset \bbR^n\) and a positive number \(\epsilon\). A subset \({\calN}_{\epsilon}\subset\calU\) is called an {\emph{\(\epsilon\)-net}} of \({}\calU\)  if every point \(\bu \in \calU\) can be approximated to within \(\epsilon\) by some point \(\bv \in \calN_{\epsilon}\), i.e., \( d(\bu,\bv) \leq \epsilon\). The {\emph{covering number}} \(\calN (\calU,\epsilon)\) is the cardinality of the smallest \(\epsilon\)-net of \(\calU\).
\end{mydef}

\begin{mydef}[Set of sparse vectors] \label{def2}
Let \(\calS^{n-1}\) be the unit sphere in \(\bbR^n\)  and \(1 \leq k \leq n\). Define
\[\calU_k \triangleq \{\bu \in \calS^{n-1}: \; \| \bu \|_0 \leq k \},
\]
also define the subset of the Euclidean  unit ball \(\calB_2^n \) with (at most) $k$-sparse vectors as
\[{\widetilde{\calU}}_k \triangleq \{ \bu \in \calB_2^{n-1}: \; \| \bu \|_0 \leq k \}.
\]
\end{mydef}
\begin{mylem} [Upper bound on covering numbers, Lemma 2.3 in \cite{UUPSubgaussian2008}] \label{lem2}
Let \(0 < \epsilon < 1\) and \(1 \leq k \leq n\). There exists an \(\epsilon\)-net of \({\tilde{\calU}}_k\), namely \({\calN_{\epsilon}}\), whose cardinality can be upper bounded as
\[\left| {\calN_{\epsilon}} \right| \leq \left(\frac{5}{2 \epsilon} \right)^k { n \choose k}.
\]
\end{mylem}
\begin{mydef}[Complexity measure \cite{UUPSubgaussian2008}] \label{def3}
The {\emph{complexity}} of a set \( \calV \subset \bbR^n \) is defined as
\[\ell_{\ast}(\calV) \triangleq {\bbE} \left[\sup_{\bv \in \calV} \left| \langle \bv, \bu \rangle \right| \right],
\]
where \(\langle \cdot ,\cdot\rangle\)   denotes inner product in $\bbR^n$, \( \bu \sim\calN(\bzero,\bI) \) is a  standard Gaussian random vector, and the supremum is over all vectors \(\bv \in \calV\).
\end{mydef}

Given a subset \(\calV \subset \bbR^n\), we aim to measure the complexity of \(\calW (\calV)\), which is the image set of the set \(\calV\) under a fixed linear mapping \(\bSigma\). More precisely, we define
\begin{equation} \label{eq:eq51}
\calW (\calV) \triangleq \{ \bw \in \bbR^n : \; \bw = \bSigma \bv, \; {\rm{for \; some}} \; \bv \in \calV \}.
\end{equation}
Define the complexity of \(\calW (\calV)\) as $\ell_{\ast} \left(\calW (\calV) \right) \triangleq \bbE  \left[\sup_{\bv \in \calV} \left| \langle \bv, \bSigma \bu \rangle \right| \right].$

\begin{mylem} [Upper bound on complexity measure, Lemma B.6 in \cite{RECSubgaussianZhou2009}] \label{lem3}
Let \({\mathcal{N}}_{\frac{1}{2}, k}\) be a \(\frac{1}{2}\)-net of \(\tilde{\calU}_k\) provided by Lemma \ref{lem2}. Then for all \(1 \leq k \leq n\), it holds that
\begin{align}
{\ell}_{\ast} \left( \calW (\calN_{\frac{1}{2},k}) \right) &\leq 3 \sqrt{k \rho_{\max}{(k)} \log{\frac{5en}{k}}}, \nonumber \\
{\ell}_{\ast} \left( \calW (\calU_k) \right) &\leq 2 {\ell}_{\ast} \left( \calW (\calN_{\frac{1}{2},k}) \right),
\end{align}
where \(\rho_{\max}{(k)}\) is the $k$-restricted maximum eigenvalue of \(\bSigma^{\ast} \bSigma\) defined in (\ref{eq:eq410}).
\end{mylem}

Define the set 
\begin{align}
  {\calE_k} \triangleq \{ \bv \in \bbR^n : \| {\bSigma} \bv \|_2 =1 ,  \| \bv \|_0 = k  \}, \label{SetcalEk}
\end{align}
then for \(\calV = \calE_k\), the complexity measure of the set \( \calW (\calE_k)\) is bounded in the following Lemma.

\begin{mylem} \label{lem4}
The complexity measure of the set \(\calW (\calE_k)\) is upper bounded as
\begin{equation} \label{eq:eq52}
{\ell}_{\ast} \left(\calW (\calE_k) \right) \leq 6 \sqrt{k {\frac{\rho_{\max} (k)}{ \rho_{\min} (k)}} \log{\frac{5en}{k}}},
\end{equation}
where \(\rho_{\max}(k)\) and \(\rho_{\min}(k)\) are defined in (\ref{eq:eq410}).
\end{mylem}

\begin{IEEEproof}
For any vector \(\bv \in \calE_k\) and any random vector \(\bu \in \bbR^n\), we have   with probability one that
\begin{align}
\left| \langle \bu, {\bSigma} \bv \rangle \right| = \left| \langle \bv, {\bSigma} \bu \rangle \right|  = \| \bv \|_2 \left| \left\langle  {\frac{\bv}{\| \bv \|_2}}, {\bSigma} \bu \right\rangle \right| & \leq  \| \bv \|_2  \sup_{\br \in \calU_k} \left| \langle  \br, {\bSigma} \bu \rangle \right|,
\label{eq:eq54}
\end{align}
where the inequality follows from  the definition of the set \(\{ \frac{\bv}{\| \bv \|_2}: \bv \in \calE_k\} \subset  \calU_k \). From Lemma \ref{lem3},
\begin{align}
{\bbE}  \left[ \sup_{\bv \in \calE_k} \left| \langle \bu, {\bSigma} \bv \rangle \right|   \right]
&\lea  \sup_{\bv \in \calE_k} \| \bv \|_2  \, \bbE \left[ \sup_{\br \in \calU_k} \left| \langle  \br, {\bSigma} \bu \rangle \right|\right] \nonumber \\
&\leb  
   6 \sqrt{k {\frac{\rho_{\max} (k)}{ \rho_{\min} (k)}} \log{\frac{5en}{k}}}.
\end{align}
where $(a)$ comes from~(\ref{eq:eq54})  and $(b)$ follows from Lemma \ref{lem3} and the definitions in (\ref{eq:eq410}).
\end{IEEEproof}

{\textbf{Step 1: Isometry on the images of sparse vectors.}} We consider the case in which the sensor data and all matrices are real. In this step, we first show that all row vectors in matrix \(\tilbZ\) are isotropic sub-Gaussian (see Definition \ref{def5} below) in Lemma \ref{lem5}. Then we use Lemma \ref{lem4} to obtain an isometry on the images of sparse vectors.

\begin{mydef}[sub-Gaussian random variables \cite{CSEldar2011}] \label{def5a}
Let \(X\) be a zero mean random variable that has unit variance. It is {\em sub-Gaussian}  if for any \(t \geq 0\), there exist a positive number \(\varrho\) such that
\[ {\bbP} \left(|X| \geq t \right) \leq 2 \exp{ \left( - \frac{t^2} {2 \varrho^2}  \right) }.
\]
The  {\em sub-Gaussian norm} \(\| X \|_{\psi_2} \) is the smallest number \(\varrho\) for which the above inequality holds.
\end{mydef}

\begin{mydef}[Isotropic sub-Gaussian random vectors \cite{CSEldar2011}] \label{def5}
Let \(\bu\) be a random vector in \(\bbR^n\). If  \( \bbE [\bu \bu^T] = \bI_n\), then \(\bu\) is called {\emph{isotropic}}. The random vector \(\bu\) is {\emph{sub-Gaussian}} with constant \(\alpha\) if
\[ \sup_{  \br \in {\bbR}^n:\|\br\|_2= 1}  {\| \langle \bu, \br \rangle \|_{\psi_2}}  < \alpha.
\]
\end{mydef}

\begin{mylem} \label{lem5}
Let \(\bu \in \bbR^n\) be a random vector with \iid \ elements, each distributed as \({\widetilde{{\calN}}} (0, 1/p, p)\). Then \(\bu\) is isotropic sub-Gaussian with constant \(\alpha = c_0 / \sqrt{p}\), where $c_0$ is an absolute constant.
\end{mylem}

\begin{IEEEproof}
Since all elements in \(\bu\) are independent zero mean random variables, and has unit variance, we have \( \bbE [\bu \bu^T] = \bI_{n} \). Let $X\sim   {{\widetilde{{\calN}}}} (0, 1 / p, p)$ be a mixed Gaussian  random variable with pdf defined in (\ref{eq:eq34}). Then, we have  for every $t\ge 0$ that
\begin{align}
\bbP ({| X |} > t ) 
&= 2 \int_{\sqrt{p} t}^{\infty} p \cdot \frac {1}{\sqrt{2 \pi}} \cdot \exp(- \frac{x^2}{2}) dx \nonumber \\
&\lea p e^{- p t^2 / 2} \leb 2 e^{- {p t^2 /2}}, \nonumber
\end{align}
where $(a)$ follows from the Chernoff bound on Gaussian $Q$-function, and $(b)$ from \( p \in (0,1]\). Hence, the sub-Gaussian norm of $X$ is bounded above by $1/\sqrt{p}$. From Lemma $5.24$ in \cite{CSEldar2011}, we have that the vector \(\bu\) is sub-Gaussian with constant \(\alpha = c_0 / \sqrt{p}\), where $c_0$ is an absolute constant.
\end{IEEEproof}

Recall that the signal model is \(\tilby = {\tilbZ} \bSigma \bx + \tilbe \). We note that all elements in matrix \(\tilbZ\) are \iid \ with distribution \( {{\widetilde{{\calN}}}} (0,1/(m p),p)\). Then Lemma \ref{lem5} implies that all row vectors of scaled matrix \( {\sqrt{m} \tilbZ}\) are independent, and isotropic sub-Gaussian with constant \(\alpha = c_0 / \sqrt{p}\). The key idea to prove Theorem \ref{mythe:the1} is to apply one result in \cite{UUPSubgaussian2008}, which is given without proof as follows.
\begin{mylem} [Theorem 2.1 in \cite{UUPSubgaussian2008}] \label{lem6}
Set \(1 \leq m \leq n\) and \(0 < \beta < 1 \). Let \(\bb\) be an isotropic sub-Gaussian random vector on \({\bbR}^n\) with constant \(\alpha \ge 1\). Let \(\bb_1,\bb_2,\dots,\bb_n\) be independent copies of \(\bb\). Let the random matrix \({\bB}\) have rows \(\bb_1,\bb_2,\dots,\bb_n\). Let \(\calV \subset \calS^{n-1}\). If \(m\) satisfies
\[m > \frac{c' \alpha^4}{\beta^2} {\ell}_{\ast} (\calV)^2,
\]
then with probability at least \(1- \exp{(- {\bar c} \beta^2 m / {\alpha^4})}\), for all \(\bv \in \calV\), we have
\[1- \beta \leq \frac{ \| { \bB} \bv \|_2^2 }{m} \leq 1+ \beta,
\]
where \(c', \bar c\) are positive absolute constants.
\end{mylem}

Recall the definitions in \eqref{eq:eq51} and \eqref{SetcalEk}, and set \( \calV = \calW(\calE_k)\). Then from Lemma \ref{lem4}, Lemma \ref{lem5} and Lemma \ref{lem6}, we obtain the following result: if the number of measurements
\begin{equation} \label{eq:eq53}
m > \frac{c_1 k \rho_{\max}(k)} {p^2 \beta^2 {\rho_{\min}(k)}} \log{\frac{5en}{k}},
\end{equation}
then with probability at least \(1- \exp{(- {c_2} \beta^2 p^2 m / {4})}\), for all \({\bv} \in \calE_k\), we have
\begin{align}
  1- \beta \leq { \| \tilbZ \bSigma \bv \|_2^2} \leq 1+ \beta,\label{eq53b}
\end{align}
where \(c_1 \triangleq 36 c' c_0^4\) and \(c_2 \triangleq \bar c / c_0^4\) are positive absolute constants.

Furthermore, by replacing \(\bf v\) with the \(\bSigma\)-normalized vector \({\bf v} / \| {\bf \Sigma}{\bf v} \|_2\) in \eqref{eq53b}, we obtain
\begin{equation} \label{eq:eq54b}
(1- \beta) \| {\bf \Sigma}{\bf v} \|_2^2  \leq { \| \tilbZ \bSigma \bv \|_2^2} \leq (1+ \beta) \| {\bf \Sigma}{\bf v} \|_2^2
\end{equation}
holds with probability at least \(1- \exp{(- {c_2} \beta^2 p^2 m / {4})}\).

{\textbf{Step 2: Restricted Isometry Property.}}
From (\ref{eq:eq54b}) and the definitions of the $k$-restricted extreme eigenvalues in (\ref{eq:eq410}), for any \(k\)-sparse vector \(\bx\), we obtain that the following inequality
\begin{equation} \label{eq:eq55}
 (1-\beta) \rho_{\min}(k) \|{\bx}\|_{2}^2 \leq { \| \tilbZ \bSigma \bx \|_2^2} \leq  (1+ \beta) \rho_{\max}(k) \|{\bx}\|_{2}^2,
\end{equation}
holds with probability at least \(1- \exp{(- {c_2} m p^2 \beta^2 / {4})}\).

Recall the definitions of the parameters \(\xi_k,\; \zeta_k,\;\vartheta_k,\;\beta_k,\) and \(\delta_k\)  defined prior to Theorem~\ref{mythe:the1}. As in (\ref{eq:eq55}), the LHS and the RHS may have different deviations from one. Hence, the maximum operation and piecewise linear mappings are used in those definitions, such that after some simple substitutions and algebraic manipulations, the following inequality
\begin{equation}\label{eq:eq56}
   (1- \delta_k) \|{\bx}\|_{2}^2 \leq { \| \tilbZ \bSigma \bx \|_2^2} \leq  (1+ \delta_k) \|{\bx}\|_{2}^2
\end{equation}
holds with probability at least \(1- \exp{(- {c_2} m p^2 \beta_k^2 / {4})} \). Collecting the results in (\ref{eq:eq53}) and (\ref{eq:eq56}), we obtain Theorem \ref{mythe:the1} for the real case.

{\textbf{Step 3: Generalization to the complex case.}} We generalize the above RIP result to the complex case. First, we show that the matrix $\tilbZ \bSigma$ satisfies the RIP for the complex data \(\bx=\bx^{\mathrm{R}} + j \bx^{\mathrm{I}}\). With probability at least \(1- \exp{(- {c_2} m p^2 \beta_k^2 / {4})} \), we have
\[
(1- \delta_k) \|\bx^{\mathrm{R}}\|_{2}^2  \leq { \| \tilbZ \bSigma \bx^{\mathrm{R}} \|_2^2}  \leq  (1+ \delta_k) \| \bx^{\mathrm{R}}\|_{2}^2,
\]
\[
(1- \delta_k) \|\bx^{\mathrm{I}}\|_{2}^2  \leq { \| \tilbZ \bSigma \bx^{\mathrm{I}} \|_2^2} \leq  (1+ \delta_k) \| \bx^{\mathrm{I}}\|_{2}^2.
\]
Combining the above two equations yields
\[
(1- \delta_k) \|\bx\|_{2}^2  \leq { \| \tilbZ \bSigma \bx \|_2^2} \leq  (1+ \delta_k) \|{\bx}\|_{2}^2.
\]

Second, we show that when the sensing matrix $\bA$ in our scheme is complex random matrix, it still satisfies the RIP. Let \(\bA = \bA^{\mathrm{R}} + j \bA^{\mathrm{I}} \). It is assumed that the real part \(\bA^{\mathrm{R}}\) and the imaginary part \(\bA^{\mathrm{I}}\) are independent, and have the same probability distribution. Recall that the sensing matrix \({\bA} = \tilbZ {\bSigma} \). For any \(k\)-sparse complex vector \(\bx\), we have
\begin{align}
  \frac{1}{2} (1- \delta_k) \|\bx\|_{2}^2  \leq {\| {\bA^{\mathrm{R}}} \bx\|_{2}^2 } \leq  \frac{1}{2} (1+ \delta_k) \|{\bx}\|_{2}^2, \nonumber \\
  \frac{1}{2} (1- \delta_k) \|\bx\|_{2}^2  \leq {\|{\bA^{\mathrm{I}}} \bx\|_{2}^2} \leq  \frac{1}{2}  (1+ \delta_k) \|{\bx}\|_{2}^2.  \nonumber
\end{align}

Combining the above two equations yields the RIP  in~\eqref{eqn:RIP} for the general complex case.
\end{IEEEproof}

\subsection{Proof of Theorem \ref{mythe:the2}} \label{SecVIIB}
\begin{IEEEproof} Clearly, we have $ \rho_{\max} (1,n)=\rho_{\min} (1,n)=1$ so the bounds are satisfied for $k=1$. We will first prove Theorem \ref{mythe:the2} for the case $k=2$. Subsequently, we generalize the result to arbitrary $2 \leq k <  \lfloor n/2 \rfloor$. Let the two non-zero elements be \(v_{s_1}= A_1 e^{j \theta_1}\) and \(v_{s_2}= A_2 e^{j \theta_2}\), where \(A_1^2 + A_2^2 = 1 \) (because $\|\bv\|_2=1$). Then from \eqref{eq:eq420}, and the fact that  ${P}_{\rm{ave}} = \sum_{i=1}^n \gamma_i / n $, we obtain
\begin{align}
  \| {\bw} \|_2^2 &= \frac{1}{n {P}_{\rm{ave}}} \sum_{i=1}^n \gamma_i + \frac{2 A_1 A_2}{n {P}_{\rm{ave}}} \sum_{i=1}^n \gamma_i \cos \left(\theta +\frac{2 \pi (i-1) \Delta}{n}\right) \nonumber  \\
&=1+2 A_1 A_2\frac{\sum_{i=1}^n a_i\gamma_i}{\sum_{i=1}^n \gamma_i},\label{SquaredNormofW}
\end{align}
where \(\theta \triangleq \theta_1 - \theta_2 \in (0, 2 \pi], \; \Delta \triangleq s_2 -s_1 \in \{1, \ldots,n-1\}, {\rm and} \; a_i \triangleq \cos (\theta +{2 \pi (i-1) \Delta} / n)\). We now set $X_i = \gamma_i$ to emphasize that the signal powers are random variables. Recall that the distributions of $X_i$'s are truncated Gaussian, denoted by $\calN_{\mathrm{tr}} (\mu, \omega)$. We consider the random variable
\begin{equation} \label{eq:eq_Sn}
S_n\triangleq\frac{\sum_{i=1}^n a_i X_i}{\sum_{i=1}^n X_i}.
\end{equation}
We define the Ces\`aro-sum of the $a_i$'s as
\begin{equation}
\bara_n \triangleq \frac{1}{n}\sum_{i=1}^n a_i = \frac{1}{n}\sum_{i=1}^n \cos \left(\theta +\frac{2 \pi (i-1) \Delta} {n} \right),\label{eqn:cesaro}
\end{equation}
and note that as $n \to \infty$, the Ces\`aro-sum converge. Indeed, we have
\begin{equation}
\bara_n \to \bara= \frac{1}{2 \pi \Delta} \int_0^{2 \pi \Delta}  \cos \left(\theta +\Delta   t  \right) dt = 0. \label{eqn:def_abar}
\end{equation}
We now bound the probability that $S_n$  exceeds some $t>0$ by considering the chain of inequalities
\begin{align}
 \bbP \left( S_n>t \right)
&=  \bbP\left( \frac{\sum_{i=1}^n a_i X_i}{\sum_{i=1}^n X_i}> t \right) \nonumber \\
&\eqa  \bbP\left(  \sum_{i=1}^n a_i X_i> t   \sum_{i=1}^n X_i\right)  \nonumber \\
&\leb  \bbP\left(   \left\{\sum_{i=1}^n a_i X_i> t    \sum_{i=1}^n X_i  \right\}\cap \left\{ \frac{1}{n}\sum_{i=1}^n X_i > \tau \mu \right\}\right)  +\bbP\left(\frac{1}{n}\sum_{i=1}^n X_i \le \tau \mu \right) \nonumber \\
&\lec  \bbP\left(  \left\{ \frac{1}{n}\sum_{i=1}^n a_i X_i>  t \tau \mu  \right\}\cap \left\{ \frac{1}{n}\sum_{i=1}^n X_i > \tau \mu \right\}\right)  +\bbP\left(\frac{1}{n}\sum_{i=1}^n X_i \le \tau \mu \right) \nonumber \\
&\le   \bbP\left(  \frac{1}{n} \sum_{i=1}^n a_i X_i>  t \tau \mu  \right)  +\bbP\left(\frac{1}{n}\sum_{i=1}^n X_i \le \tau \mu \right), \label{eqn:two_probs}
\end{align}
where $(a)$ is due to the fact that $X_i$'s are nonnegative random variables, (b) follows from the fact $\bbP(\calA)=\bbP(\calA\cap\calB)+\bbP(\calA\cap\calB^c)\le \bbP(\calA\cap\calB)+\bbP(\calB^c)$ and $(c)$ comes from monotonicity of measure. In the following, we bound the two terms in \eqref{eqn:two_probs} using  the theory of large deviations~\cite{LargeDeviation1998}. 

Define $t'\triangleq t \tau \mu$ and let $s$ be an arbitrary non-negative number. Then from Markov's inequality, the first term in \eqref{eqn:two_probs} can be upper bounded as follows
\begin{align}
\bbP\left(  \frac{1}{n} \sum_{i=1}^n a_i X_i>  t' \right)\le\exp(-nst')\bbE \left[ \exp\left(\sum_{i=1}^n s a_i X_i \right)\right],
\end{align}
which implies by the independence of  the $X_i$'s that
\begin{align}
\frac{1}{n}\log\bbP\left(  \frac{1}{n} \sum_{i=1}^n a_i X_i >  t' \right)\le -st'+ \frac{1}{n}\sum_{i=1}^n\log\bbE [ \exp(s a_i X_i) ] \label{eqn:log_prob}.
\end{align}
To bound the sum in \eqref{eqn:log_prob}, we find the  cumulant-generating function (CGF) of $X\sim \calN_{\mathrm{tr}}(\mu, \omega^2)$ in terms of a Gaussian with mean $\mu$ and variance $\omega^2$. By simple algebraic manipulations, we have
\begin{align}
\log \bbE [ \exp(s X) ]= \mu s + \frac{1}{2} \omega^2 s^2 + \varphi (\mu, \omega, s), \label{eq:CGF}
\end{align}
where $\varphi (\mu, \omega, s) \triangleq\log \left( 1 - Q\left({\mu}/{\omega} + \omega s \right) \right) - \log \left(1 - Q\left({\mu}/{\omega} \right) \right) $. We note that given that $(\mu,\omega)$ is a positive pair of numbers,  $s\mapsto\varphi (\mu, \omega, s)$ for $s\ge 0$ is concave, because $s \mapsto - Q(\mu/\omega+\omega s)$   (for $\mu/\omega>0$)  and $t\mapsto\log(1+t)$ are both concave  and the latter function is non-decreasing. Moreover, $s\mapsto\varphi (\mu, \omega, s)$ is continuous for each positive $(\mu, \omega)$ pair, because  every concave function on an open set is continuous. Note that $\varphi (\mu, \omega,0)=0$.

Substituting  the CGF of the truncated Gaussian distribution in \eqref{eq:CGF} into \eqref{eqn:log_prob} yields
\begin{align}
&\frac{1}{n}\log\bbP\left(  \frac{1}{n} \sum_{i=1}^n a_i X_i>  t' \right) \nonumber \\
& \le -st'+ {\mu s} \bara_n + \frac{\omega^2 s^2}{2n} \sum_{i=1}^n  a_i^2 + \frac{1}{n} \sum_{i=1}^n \varphi (\mu, \omega, a_i s)  \nonumber \\
&\eqa -st'+ {\mu s} \bara_n +  \frac{\omega^2 s^2}{4} + \frac{\omega^2 s^2}{4n} \sum_{i=1}^n  \cos \left( 2 \theta + \frac{4 \pi \Delta (i-1)}{n}\right) + \frac{1}{n} \sum_{i=1}^n \varphi (\mu, \omega, a_i s)  \nonumber   \\
&\leb  -st'+ {\mu s} \bara_n +  \frac{\omega^2 s^2}{4} + \frac{\omega^2 s^2}{4n} \sum_{i=1}^n  \cos \left( 2 \theta + \frac{4 \pi \Delta (i-1)}{n}\right) + \varphi \left(\mu, \omega, \frac{s}{n}\sum_{i=1}^n a_i \right), \label{eqn:log_prob_simp}
\end{align}
where $(a)$ comes from the definition of $a_i$ and the double-angle formula for the cosine, and $(b)$ follows the fact $\varphi(\mu, \omega, s)$ is concave in $s$ for any positive $(\mu,\omega)$ pair.

Taking the limsup on both sides of \eqref{eqn:log_prob_simp} and using the definition of $\bara_n$ yields
\begin{align}
&\limsup_{n \to \infty} \frac{1}{n}\log\bbP\left(  \frac{1}{n} \sum_{i=1}^n a_i X_i>  t' \right)  \nonumber \\
& \lea -st'+  \frac{\omega^2 s^2}{4} + \frac{\omega^2 s^2}{16 \pi \Delta} \int_0^{4 \pi \Delta} \cos \left( 2 \theta + t\right) dt + \limsup_{n \to \infty} \varphi \left(\mu, \omega, \bara_n s\right) \nonumber \\ 
& \eqb -st'+  \frac{\omega^2 s^2}{4}  + \limsup_{n \to \infty} \varphi \left(\mu, \omega, \bara_n s\right) \nonumber \\ 
&\eqc -st'+ \frac{\omega^2 s^2}{4} \triangleq f(s), \label{eqn:def_f}
\end{align}
where (a)  follows from Riemann sums, (b)   comes from the fact cosine has zero mean over an integer number of periods (note $\Delta\in\bbZ$) and (c) follows from  the continuity of $\varphi (\mu, \omega, s)$ and \eqref{eqn:def_abar}. Note that  the minimum  $f(s)$ in~\eqref{eqn:def_f} is $f(s^{\ast}) = - \tau^2 d^2 t^2$ (attained at $s^{\ast} = 2t'/\omega^2$). Hence, 
\begin{align}
 \bbP\left(  \frac{1}{n} \sum_{i=1}^n a_i X_i>  t' \right) \dotleq \exp\left[ - n \tau^2 d^2 t^2 \right]. \label{eqn:log_prob_simp_limisup}
\end{align}

The second term in \eqref{eqn:two_probs} can be bounded using standard techniques from the large deviations theory \cite{LargeDeviation1998} (Cram\'er's theorem) and along the same lines as the derivation above. As such we have
\begin{align}
 \bbP\left(\frac{1}{n}\sum_{i=1}^n X_i \le \tau  \mu \right) & \lea \exp\left[ -n \left( s\mu(1-\tau)-\omega^2 s^2/2-\varphi(\mu,\omega,-s)\right)\right] \nonumber \\
& \leb \exp\left[ -n \left( s\mu(1-\tau)-\omega^2 s^2/2 \right)\right], \nonumber
\end{align}
where $(a)$ follows from using the CGF of $X_i$ in~\eqref{eq:CGF}, and $(b)$ follows from the fact that $\varphi(\mu,\omega,-s)\le 0$ for all $s\ge 0$. Hence, setting  $s\triangleq\mu(1-\tau)/\omega^2$, we have
\begin{equation}
 \bbP\left(\frac{1}{n}\sum_{i=1}^n X_i \le \tau  \mu \right)\dotleq\exp\left[ -n(1-\tau)^2d^2/2\right]. \label{eqn:prob_2_limsup}
\end{equation}

Combining the two terms in \eqref{eqn:two_probs}, we have from \eqref{eqn:log_prob_simp_limisup} and \eqref{eqn:prob_2_limsup} and the largest-exponent-dominates principle that
\begin{equation}
\bbP(S_n>t )\dotleq\exp\left[ -n \min\left\{ \tau^2 d^2 t^2, (1-\tau)^2 d^2 / 2 \right\}\right] \label{eqn:log_prob_comb1_limisup}
\end{equation}

Since $\tau>0$ is  a free parameter, we can set it to be $ \tau^{\ast}\triangleq\frac{1}{1+\sqrt{2} t}$. Substituting $\tau^*$ into \eqref{eqn:log_prob_comb1_limisup} yields 
\begin{align}
  \bbP \left(S_n>t\right)   \dotleq - n d^2 {\tilde t}^2. \label{eqn:log_prob_comb2_limisup}
\end{align}
where ${\tilde t} \triangleq t / (1+\sqrt{2} t)$. By symmetry, we can also conclude that
\begin{align}
\bbP \left(S_n<-t\right)   \dotleq - n d^2 {\tilde t}^2. \label{eqn:log_prob_comb2_limisup1}
\end{align}

Recall  that $\rho_{\max}(k,n)$  is the maximum value of  $\| {\bw} \|_2^2 = \|\bSigma \bv \|_2^2$ over  all unit-norm $k$-sparse vectors $\bv$. From \eqref{SquaredNormofW}, $\| {\bw} \|_2^2$ depends only on   $A_1 A_2$. Note that \(0 < A_1 A_2 \leq 1/2\)    because  \(\sqrt{A_1 A_2} \leq (A_1 + A_2) / 2 \). We set \(A_1 A_2 = 1/2\), whence  $\|\bw\|_2$ attains its maximum value. From \eqref{SquaredNormofW},
\begin{equation} \label{eq:eqrho_limsup}
\begin{split}
 \bbP \left( \rho_{\max} (2,n) > 1+ t \right) &\dotleq \exp\left[- n d^2 {\tilde t}^2 \right],  \\
 \bbP \left( \rho_{\min} (2,n) < 1 -  t \right) &\dotleq \exp\left[- n d^2 {\tilde t}^2 \right].
\end{split}
\end{equation}

Having proved the result for the $k=2$ case, we now generalize it to the case where $k > 2$. Set the non-zero elements of the vector $\bv$ to be  \(v_{s_q}= A_q e^{j \theta_q}, \; q=1, \ldots ,k\), where \(\sum \nolimits_{q=1}^k A_q^2 = 1\). Equation \eqref{eq:eq420} can be written as
\begin{align}
\| {\bw} \|_2^2 &= \frac{1}{n {P}_{\rm{ave}}} \sum_{i=1}^n \gamma_i  \left(  1 +  \sum_{q=1}^k  \sum_{l=1, l \neq q}^k  A_q A_l \cos \left( \theta_{q,l} + \frac{j 2 \pi (i-1) \Delta_{q,l}}{n} \right) \right) \nonumber \\
&= 1 +  \sum_{q=1}^k  \sum_{l=1, l \neq q}^k  A_q A_l \sum_{i=1}^k \cos \left(\theta_{q,l} + \frac{2 \pi (i-1) \Delta_{q,l}}{n}\right), \nonumber \\
&= 1 +  \sum_{q=1}^k  \sum_{l=1, l \neq q}^k  A_q A_l S_n^{q,l} \triangleq 1+B_n, \label{eq:eq424b}
\end{align}
where $S_n^{q,l}$ is defined as in \eqref{eq:eq_Sn} but involving the $q$-th and the $l$-th nonzero elements of $\bv$, i.e., $\theta_{q,l}=\theta_q- \theta_l$, and $\Delta_{q,l}=s_l - s_q$. On the other hand, we can bound $B_n^2$ as follows
\begin{align}
B_n^2 &\lea \left(\sum_{q=1}^k  \sum_{l=1, l \neq q}^k  A_q^2 A_l^2 \right) \left(\sum_{q=1}^k  \sum_{l=1, l \neq q}^k (S_n^{q,l})^2\right) \nonumber \\
&= \left( \left(\sum_{q=1}^k  A_q^2\right)^2 - \sum_{q=1}^k  A_q^4 \right) \left(\sum_{q=1}^k  \sum_{l=1, l \neq q}^k (S_n^{q,l})^2\right) \nonumber \\
&\leb \left( 1 - \frac{1}{k} \left(\sum_{q=1}^k  A_q^2\right)^2 \right) \left(\sum_{q=1}^k  \sum_{l=1, l \neq q}^k (S_n^{q,l})^2\right) \nonumber \\
&= \frac{k-1}{k} \sum_{q=1}^k  \sum_{l=1, l \neq q}^k (S_n^{q,l})^2, \label{eq:B_n_upperbound}
\end{align}
where (a) comes from the Cauchy-Schwartz inequality and (b) comes from the basic inequality relating the  arithmetic  and quadratic means, namely $1/M\sum_{j=1}^M\alpha_j\le  (1/M\sum_{j=1}^M\alpha_j^2)^{1/2}$.

Now, given any $t >0$, we can   bound the probability that $|B_n|$ exceeds $t$ as follows:
\begin{align}
\bbP \left( | B_n | > t\right) &= \bbP \left( B_n^2> t^2\right) \nonumber \\
&\lea \bbP \left(\sum_{q=1}^k  \sum_{l=1, l \neq q}^k (S_n^{q,l})^2   >  \frac{k t^2 }{k-1}\right) \nonumber \\
&\le  \bbP \left(\max_{l \neq q} (S_n^{q,l})^2  >  \frac{t^2 }{(k-1)^2} \right) \nonumber \\
&\leb \sum_{q=1}^k  \sum_{l=1, l \neq q}^k \bbP \left(\left( S_n^{q,l}\right)^2  >  \frac{t^2 }{(k-1)^2}\right) \nonumber \\
&= \sum_{q=1}^k  \sum_{l=1, l \neq q}^k \bbP \left( S_n^{q,l}  >  \frac{t }{k-1}\right) , \label{eq:B_n_tail}
\end{align}
where (a) comes from \eqref{eq:B_n_upperbound} and monotonicity of measure and (b) comes from the union bound. Applying the result for $k=2$ in~\eqref{eq:eqrho_limsup} to \eqref{eq:B_n_tail}, we have 
\begin{align}
\bbP \left( | B_n | > t\right) \dotleq k(k-1) \exp \left[- n d^2 E(k,t)^2   \right], \label{eq:B_n_tail_exp}
\end{align}
where the exponent is $E(k,t)\triangleq t /  (k-1+\sqrt{2} t)$. Recall  the definition of $\rho_{\max}(k,n)$ in \eqref{eq:eq410}. From \eqref{eq:eq424b} and~\eqref{eq:B_n_tail_exp}, we conclude that
\begin{align}
\bbP \left( \rho_{\max} (k,n) > 1+ t \right) \dotleq\exp   \left[- n d^2 E(k,t)^2   \right].
\end{align}
The analysis of $\bbP \left( \rho_{\min} (k,n) < 1- t \right)$ proceeds {\em mutatis mutandis}. This completes the proof.
\end{IEEEproof}



\renewcommand{\baselinestretch}{1.45}

\bibliography{IEEEabrv,reference2}
\bibliographystyle{ieeetr}

\end{document}

%% file: preamble3.tex
\usepackage{cite}
\usepackage[cmex10]{amsmath}
\usepackage{amssymb}
\usepackage{amsthm}
\usepackage{mathrsfs}
\usepackage{bm}
\usepackage[mathscr]{eucal}
\usepackage{amssymb,amsmath,amsthm,amsfonts,latexsym}
\usepackage{amsmath,graphicx,bm,xcolor,url}
\usepackage{graphicx}
\usepackage{latexsym}
\usepackage{CJK}
\usepackage{indentfirst}
\usepackage{geometry}
\usepackage{psfrag}
\usepackage{setspace}
\usepackage{algorithmic}
\usepackage{algorithmic, cite}
\usepackage{algorithm}
\usepackage{array}
\usepackage{mdwmath}
\usepackage{mdwtab}
\usepackage{eqparbox}
\usepackage[tight,footnotesize]{subfigure}
\usepackage{url}
\usepackage{epstopdf}
\usepackage{epsfig,epsf,psfrag}
\usepackage{fixltx2e}
\usepackage{verbatim}
\usepackage{textcomp}
\usepackage{psfrag} 

\usepackage{subfigure} 

\theoremstyle{plain}
\newtheorem{mythe}{Theorem}
\theoremstyle{remark}
\newtheorem{mylem}{Lemma}
\theoremstyle{plain}
\newtheorem{mydef}{Definition}
\theoremstyle{remark}

\theoremstyle{plain}
\newtheorem{mycor}{Corollary}
\theoremstyle{remark}
\newtheorem{myrem}{Remark}
\theoremstyle{remark}
\newtheorem{myexa}{Example}
\theoremstyle{remark}

\theoremstyle{remark}

\theoremstyle{remark}

\theoremstyle{remark}

\catcode`~=11 \def\UrlSpecials{\do\~{\kern -.15em\lower .7ex\hbox{~}\kern .04em}} \catcode`~=13

\allowdisplaybreaks[4]

\newcommand{\calA}{\mathcal{A}}
\newcommand{\calB}{\mathcal{B}}

\newcommand{\calE}{\mathcal{E}}

\newcommand{\calN}{\mathcal{N}}

\newcommand{\calS}{\mathcal{S}}
\newcommand{\calT}{\mathcal{T}}
\newcommand{\calU}{\mathcal{U}}
\newcommand{\calV}{\mathcal{V}}
\newcommand{\calW}{\mathcal{W}}

\newcommand{\ba}{\mathbf{a}}
\newcommand{\bA}{\mathbf{A}}
\newcommand{\bb}{\mathbf{b}}
\newcommand{\bB}{\mathbf{B}}

\newcommand{\be}{\mathbf{e}}

\newcommand{\bH}{\mathbf{H}}

\newcommand{\bI}{\mathbf{I}}

\newcommand{\br}{\mathbf{r}}

\newcommand{\bs}{\mathbf{s}}

\newcommand{\bu}{\mathbf{u}}

\newcommand{\bv}{\mathbf{v}}

\newcommand{\bw}{\mathbf{w}}

\newcommand{\bx}{\mathbf{x}}

\newcommand{\by}{\mathbf{y}}

\newcommand{\bZ}{\mathbf{Z}}

\newcommand{\rme}{\mathrm{e}}

\newcommand{\rmH}{\mathrm{H}}

\newcommand{\rmR}{\mathrm{R}}

\newcommand{\bbC}{\mathbb{C}}

\newcommand{\bbE}{\mathbb{E}}

\newcommand{\bbP}{\mathbb{P}}

\newcommand{\bbR}{\mathbb{R}}

\newcommand{\bbZ}{\mathbb{Z}}

\DeclareMathAlphabet{\mathbsf}{OT1}{cmss}{bx}{n}
\DeclareMathAlphabet{\mathssf}{OT1}{cmss}{m}{sl}

\DeclareSymbolFont{bsfletters}{OT1}{cmss}{bx}{n}
\DeclareSymbolFont{ssfletters}{OT1}{cmss}{m}{n}
\DeclareMathSymbol{\bsfGamma}{0}{bsfletters}{'000}
\DeclareMathSymbol{\ssfGamma}{0}{ssfletters}{'000}
\DeclareMathSymbol{\bsfDelta}{0}{bsfletters}{'001}
\DeclareMathSymbol{\ssfDelta}{0}{ssfletters}{'001}
\DeclareMathSymbol{\bsfTheta}{0}{bsfletters}{'002}
\DeclareMathSymbol{\ssfTheta}{0}{ssfletters}{'002}
\DeclareMathSymbol{\bsfLambda}{0}{bsfletters}{'003}
\DeclareMathSymbol{\ssfLambda}{0}{ssfletters}{'003}
\DeclareMathSymbol{\bsfXi}{0}{bsfletters}{'004}
\DeclareMathSymbol{\ssfXi}{0}{ssfletters}{'004}
\DeclareMathSymbol{\bsfPi}{0}{bsfletters}{'005}
\DeclareMathSymbol{\ssfPi}{0}{ssfletters}{'005}
\DeclareMathSymbol{\bsfSigma}{0}{bsfletters}{'006}
\DeclareMathSymbol{\ssfSigma}{0}{ssfletters}{'006}
\DeclareMathSymbol{\bsfUpsilon}{0}{bsfletters}{'007}
\DeclareMathSymbol{\ssfUpsilon}{0}{ssfletters}{'007}
\DeclareMathSymbol{\bsfPhi}{0}{bsfletters}{'010}
\DeclareMathSymbol{\ssfPhi}{0}{ssfletters}{'010}
\DeclareMathSymbol{\bsfPsi}{0}{bsfletters}{'011}
\DeclareMathSymbol{\ssfPsi}{0}{ssfletters}{'011}
\DeclareMathSymbol{\bsfOmega}{0}{bsfletters}{'012}
\DeclareMathSymbol{\ssfOmega}{0}{ssfletters}{'012}

\newcommand{\tile}{\widetilde{e}}

\newcommand{\tilbe}{\widetilde{\be}}

\newcommand{\tilbH}{\widetilde{\bH}}

\newcommand{\hatbs}{\widehat{\bs}}

\newcommand{\hatbx}{\widehat{\bx}}

\newcommand{\tilby}{\widetilde{\by}}

\newcommand{\tilZ}{\widetilde{Z}}

\newcommand{\tilbZ}{\widetilde{\bZ}}

\newcommand{\bara}{\bar{a}}

\newcommand{\bpsi}{\bm{\psi}}

\newcommand{\bGamma}{\bm{\Gamma}}

\newcommand{\bSigma	}{\bm{\Sigma}}
\newcommand{\bPsi}{\bm{\Psi}}

\newcommand{\bPhi}{\bm{\Phi}}

\newcommand{\tsigma}{\widetilde{\sigma}}

\newcommand{\tbPhi}{\widetilde{\bPhi}}

\newcommand{\iid}{i.i.d.}

\def\norm#1{\left\| #1 \right\|}
\def\norm2#1{\left\| #1 \right\|_2}
\def\norm22#1{\left\| #1 \right\|_2^2}

\newcommand{\dotleq}{\stackrel{.}{\leq}}

\newcommand{\eqa}{\stackrel{(a)}{=}}
\newcommand{\eqb}{\stackrel{(b)}{=}}
\newcommand{\eqc}{\stackrel{(c)}{=}}

\newcommand{\lea}{\stackrel{(a)}{\le}}
\newcommand{\leb}{\stackrel{(b)}{\le}}
\newcommand{\lec}{\stackrel{(c)}{\le}}

\DeclareMathOperator{\st}{subject\,\,to}

\DeclareMathOperator{\diag}{diag}

\newcommand{\bzero}{\mathbf{0}}

\newcommand{\qednew}{\nobreak \ifvmode \relax \else
      \ifdim\lastskip<1.5em \hskip-\lastskip
      \hskip1.5em plus0em minus0.5em \fi \nobreak
      \vrule height0.75em width0.5em depth0.25em\fi}

%% file: mac2.tex
\begin{figure}[t]
\centering
\setlength{\unitlength}{.4mm}
\begin{picture}(292,160)
\thicklines
\put(0, 0){\line(1,0){20}}
\put(0, 150){\line(1,0){20}}
\put(0, 0){\line(0,1){150}}
\put(20, 0){\line(0,1){150}}

\put(20, 140){\vector(1,0){40}}
\put(20, 90){\vector(1,0){40}}
\put(20, 10){\vector(1,0){40}}

\put(68, 140){\circle{16}}
\put(68, 90){\circle{16}}
\put(68, 10){\circle{16}}

\put(61, 137){  $\times$}
\put(61, 87){  $\times$}
\put(61, 7){  $\times$}

\put(68, 168){\vector(0,-1){20}}
\put(68, 118){\vector(0,-1){20}}
\put(68, 38){\vector(0,-1){20}}

\put(4, 73){  $\bs$}

\put(26, 144){  $\phi_{i1} s_1$}
\put(26, 94){  $\phi_{i2} s_2$}
\put(26, 14){  $\phi_{in} s_n$}

\put(37, 47){  $\vdots$}

\put(50, 161){  $h_{i1}$}
\put(50, 111){  $h_{i2}$}
\put(50, 31){  $h_{in}$}

\put(116, 37){\line(1,0){20}}
\put(116, 112){\line(1,0){20}}
\put(116, 37){\line(0,1){75}}
\put(136, 37){\line(0,1){75}}

\put(121, 70){$\sum$}

\put(76, 140){\vector(1,-1){40}}
\put(76, 90){\vector(1,0){40}}
\put(76, 10){\vector(1,1){40}}

%


%
%
%

\put(136, 75){\vector(1,0){36}}
\put(200, 64){  $y_i$}
\put(182, 75){\circle{16}}
\put(182, 103){\vector(0,-1){20}}
\put(175, 72){  $+$}
\put(168, 94){  $e_i$}

\put(192, 75){\vector(1,0){30}}

\put(222, 45){\line(1,0){40}}
\put(222, 105){\line(1,0){40}}
\put(222, 45){\line(0,1){60}}
\put(262, 45){\line(0,1){60}}

\put(227, 77){Fusion}
\put(227, 63){Center}

\end{picture}
\caption{The MAC  communication structure for WSNs in the $i$-th time slot. The signals that are concurrently transmitted from sensors to the FC  are linearly combined over the air.}
\label{fig:Fig1}
\end{figure} 

%% file: Draft_YG_22Oct_Final.bbl
\begin{thebibliography}{10}

\bibitem{Kansal}
A.~Kansal, J.~Hsu, S.~Zahedi, and M.~B. Srivastava, ``Power management in
  energy harvesting sensor networks,'' {\em ACM Trans. Embed. Comput. Syst.},
  vol.~6, Sept. 2007.

\bibitem{CKTSP2012}
C.~K. Ho and R.~Zhang, ``Optimal energy allocation for wireless communications
  with energy harvesting constraints,'' {\em {IEEE} Trans. Signal Process.},
  vol.~60, pp.~4808--4818, Sept. 2012.

\bibitem{IntroductionCS06}
E.~J. Candes and M.~B. Wakin, ``An introduction to compressive sampling,'' {\em
  {IEEE} Signal Process. Mag.}, vol.~25, pp.~21--30, Mar. 2008.

\bibitem{ReconRandProjHaupt2006}
J.~D. Haupt and R.~D. Nowak, ``Signal reconstruction from noisy random
  projections,'' {\em {IEEE} Trans. Inf. Theory}, vol.~52, pp.~4036--4048,
  Sept. 2006.

\bibitem{InfBoundAeron2007}
S.~Aeron, M.~Zhao, and V.~Saligrama, ``Information theoretic bounds to sensing
  capacity of sensor networks under fixed {SNR},'' in {\em IEEE Inf. Th.
  Workshop}, (Lake Tahoe, CA, USA), pp.~84--89, Sept. 2007.

\bibitem{EASTRana2010}
R.~Rana, W.~Hu, and C.~T. Chou, ``Energy-aware sparse approximation technique
  ({EAST}) for rechargeable wireless sensor networks,'' in {\em European Conf.
  on Wireless Sensor Networks}, (Coimbra, Portugal), pp.~306--321, Feb. 2010.

\bibitem{XueICC2012}
T.~Xue, X.~Dong, and Y.~Shi, ``A multiple access scheme based on
  multi-dimensional compressed sensing,'' in {\em IEEE Int. Conf. on Commun.
  (ICC)}, (Ottawa, Canada), pp.~3832--3836, Jun. 2012.

\bibitem{Fazel2011}
F.~Fazel, M.~Fazel, and M.~Stojanovic, ``Random access compressed sensing for
  energy-efficient underwater sensor networks,'' {\em {IEEE} J. Sel. Areas
  Commun.}, vol.~29, pp.~1660--1670, Sept. 2011.

\bibitem{PhDThesisBajwa2009}
W.~Bajwa, {\em New information processing theory and methods for exploiting
  sparsity in wireless systems}.
\newblock PhD thesis, University of Wisconsin-Madison, 2009.

\bibitem{RMTinWC2004}
A.~M. Tulino and S.~Verd\'{u}, {\em Random matrix theory and wireless
  communications}.
\newblock Hanover, MA, USA: now Publishers Inc. Press, 2004.

\bibitem{DLPCandes05}
E.~J. Candes and T.~Tao, ``Decoding by linear programming,'' {\em {IEEE} Trans.
  Inf. Theory}, vol.~51, pp.~4203--4215, Dec. 2005.

\bibitem{SimpleRIP2008}
R.~Baraniuk, M.~Davenport, R.~D. Vore, and M.~Wakin, ``A simple proof of the
  restricted isometry property,'' {\em Constr. Approx.}, vol.~28, no.~3,
  pp.~253--263, 2008.

\bibitem{CSEldar2011}
Y.~C. Eldar and G.~Kutyniok, {\em Compressed sensing: {Theory} and
  applications}.
\newblock Cambridge Univ. Press, 2012.

\bibitem{UUPSubgaussian2008}
S.~Mendelson, A.~Pajor, and N.~T. Jaegermann, ``Uniform uncertainty principle
  for {Bernoulli} and {sub-Gaussian} ensembles,'' {\em Constr. Approx.},
  vol.~28, pp.~277--289, 2008.

\bibitem{NewRICCai}
T.~T. Cai, M.~Wang, and G.~Xu, ``New bounds for restricted isometry
  constants,'' {\em {IEEE} Trans. Inf. Theory}, vol.~56, pp.~4388--4394, Sept.
  2010.

\bibitem{DavenportNoisefold11}
M.~A. Davenport, ``The pros and cons of compressive sensing for wideband signal
  acquisition: {N}oise folding versus dynamic range,'' {\em {IEEE} Trans.
  Signal Process.}, vol.~60, pp.~4628--4642, Sept. 2012.

\bibitem{LargeDeviation1998}
A.~S. Dembo and O.~Zeitouni, {\em Large deviation techniques and applications}.
\newblock Springer Press, 1998.

\bibitem{BPDNBerg}
E.~V.~D. Berg and M.~P. Friedlander, ``Probing the {P}areto frontier for basis
  pursuit solutions,'' {\em Proc. of Soc. Ind. Appl. Math.}, vol.~31, no.~2,
  pp.~890--912, 2008.

\bibitem{RECSubgaussianZhou2009}
S.~Zhou, ``Restricted eigenvalue conditions on {sub-Gaussian} random
  matrices.'' Website, 2009.
\newblock \url{http://arxiv.org/abs/0912.4045v2}.

\end{thebibliography}
